\documentclass[aps, prd, amsmath, amssymb, floats, floatfix, twocolumn, nofootinbib, superscriptaddress]{revtex4-1}
\usepackage{graphicx}
\usepackage{color}
\usepackage{latexsym}
\usepackage{epsfig}
\usepackage[colorlinks]{hyperref}
\usepackage{array}
\usepackage{enumitem}
\usepackage{textcomp}
\usepackage{footmisc}
\setlist{noitemsep,leftmargin=*}

\newcommand{\be}{\begin{eqnarray}}
\newcommand{\ee}{\end{eqnarray}}
\newcommand{\rar}{\rightarrow}
\newcommand{\bitem}{\begin{itemize}}
\newcommand{\eitem}{\end{itemize}}
\newcommand{\bwide}{\begin{widetext}}
\newcommand{\ewide}{\end{widetext}}

\newcommand{\ch}[1]{\textcolor{black}{{#1}}}

\begin{document}

\title{Testing General Relativity with X-ray reflection spectroscopy: The Konoplya-Rezzolla-Zhidenko parametrization}

\author{Sourabh~Nampalliwar}
\email[Corresponding author: ]{sourabh.nampalliwar@uni-tuebingen.de}
\affiliation{Theoretical Astrophysics, Eberhard-Karls Universit\"at T\"ubingen, 72076 T\"ubingen, Germany}

\author{Shuo~Xin}
\affiliation{School of Physics Sciences and Engineering, Tongji University, Shanghai 200092, China}

\author{Shubham~Srivastava}
\affiliation{Indian Institute of Technology, Kharagpur 721302, India}

\author{Askar~B.~Abdikamalov}
\affiliation{Center for Field Theory and Particle Physics and Department of Physics, Fudan University, 200438 Shanghai, China}

\author{Dimitry~Ayzenberg}
\affiliation{Center for Field Theory and Particle Physics and Department of Physics, Fudan University, 200438 Shanghai, China}
\affiliation{Theoretical Astrophysics, Eberhard-Karls Universit\"at T\"ubingen, 72076 T\"ubingen, Germany}

\author{Cosimo~Bambi}
\affiliation{Center for Field Theory and Particle Physics and Department of Physics, Fudan University, 200438 Shanghai, China}

\author{Thomas~Dauser}
\affiliation{Remeis Observatory \& ECAP, Universit\"{a}t Erlangen-N\"{u}rnberg, 96049 Bamberg, Germany}

\author{Javier~A.~Garc{\'\i}a}
\affiliation{Cahill Center for Astronomy and Astrophysics, California Institute of Technology, Pasadena, CA 91125, USA}
\affiliation{Remeis Observatory \& ECAP, Universit\"{a}t Erlangen-N\"{u}rnberg, 96049 Bamberg, Germany}

\author{Ashutosh~Tripathi}
\affiliation{Center for Field Theory and Particle Physics and Department of Physics, Fudan University, 200438 Shanghai, China}

\begin{abstract}
X-ray reflection spectroscopy is a promising technique for testing general relativity in the strong field regime, as it can be used to test the Kerr black hole hypothesis. In this context, the parametrically deformed black hole metrics proposed by Konoplya, Rezzolla \& Zhidenko (Phys. Rev. D93, 064015, 2016) form an important class of non-Kerr black holes. We implement this class of black hole metrics in \textsc{relxill\_nk}, which is a framework we have developed for testing for non-Kerr black holes using X-ray reflection spectroscopy. \ch{We perform a qualitative analysis of the effect of the leading order strong-field deformation parameters on typical observables like the innermost stable circular orbits and the reflection spectra.}
\ch{We also present the first X-ray constraints on some of the deformation parameters of this metric, using \textit{Suzaku} data from the supermassive black hole in Ark~564, and compare them with those obtained (or expected) from other observational techniques like gravitational waves and black hole imaging.}
\end{abstract}

\maketitle


\section{Introduction \label{s-intro}}

Einstein's theory of gravity, since its proposition over a century ago, has been applied to a variety of astrophysical phenomena in our Universe. Over these years, it has emerged as the standard framework for describing the spacetime in the presence of gravitational objects. While largely successful in the weak-field tests~\cite{will2014}, only recently the strong-field predictions of Einstein's gravity (the general theory of relativity, GR hereafter) have become testable in a variety of ways~\cite{Bambi2015,LIGOScientific:2019fpa,Akiyama:2019fyp}. Presence of a zoo of \textit{alternative theories of gravity}, which address shortcomings of GR with respect to observations e.g., dark matter and dark energy, and/or extend GR to overcome issues e.g., difficulties in quantizing GR and resolution of the curvature singularity, make it crucial to test the strong-field predictions of GR with the latest techniques and technologies.

Black holes (BH hereafter) are surprisingly ubiquitous objects in our Universe, and due to strong gravity regions in their neighborhoods, form the perfect candidates for testing theories of gravity. Within GR, under typical astrophysical conditions, a BH is an extremely simple object and its effect on the spacetime is described by very few parameters. Most commonly, these are the BH \textit{mass} and \textit{spin} and the object is known as a Kerr BH~\cite{Kerr1963}. The assumption that astrophysical BHs are described by the Kerr metric is known as the \textit{Kerr hypothesis}. (For the specific conditions and assumptions, see Ref.~\cite{Chrusciel2012}.) Alternative theories of gravity often introduce additional parameters, deforming the BH away from the Kerr solution. Observations of effects of BH have been a celebrated exercise in physics, given the potential for discovery of interesting phenomena. Some of the ways these observations are done are: X-ray spectroscopy (first measurements of the BH spin~\cite{Dabrowski:1997xr,Young:1998ha}), gravitational wave interferometry (first observation of coalescence of a pair of BHs~\cite{Abbott:2016blz}), pulsar timing (first indirect detection of gravitational waves~\cite{Taylor:1989sw}) and BH imaging (capturing an \textit{image} of the region close to the BH horizon for the first time~\cite{Akiyama:2019cqa}).

In this work, our focus is on the technique of X-ray spectroscopy. In particular, we are interested in the reflection spectrum of BHs with accretion disks, which is in the X-ray band. Since the gravity of the BH affects its neighborhood and photons that travel from the neighborhood to us, the analysis of the reflection spectrum can be used to study the nature of BH itself~\cite{Bambi2015,Bambi2018}. The most advanced model for calculation of the reflection spectrum in the Kerr case is \textsc{relxill}~\cite{Garcia2013, Dauser2014}. We have extended this model to \textsc{relxill\_nk}, which can calculate the reflection spectrum for non-Kerr metrics~\cite{relxillnk,Abdikamalov:2019yrr}. The model has been applied to X-ray observations of several astrophysical BHs to place constraints on deviations away from the Kerr solution~\cite{Cao2017,Tripathi2018a,Wang-Ji:2018ssh,Xu2018,Choudhury:2018zmf,Zhou2018a,Zhou2018b,Tripathi:2018lhx,Tripathi:2019bya,Zhang:2019zsn,Tripathi:2019fms}. 

The non-Kerr metric most commonly used with \textsc{relxill\_nk} is the one proposed in Ref.~\cite{Johannsen2015} (See also~\cite{Vigeland2011}). This metric, referred to as the Johannsen metric hereafter, albeit not a solution of a well-defined alternative theory of gravity, \ch{preserves the symmetry of the Kerr metric associated with a Killing tensor~\cite{Carter:1968rr,Papadopoulos:2018nvd}. It can be mapped to BH solutions from some alternative theories with suitable combinations of the new parameters introduced in the metric. In this sense, it is a good candidate for theory-agnostic tests of the Kerr hypothesis. But it suffers from some important limitations. BHs in alternative theories do not generally have the additional symmetry of the Kerr solution. Moreover, the parametrization used in Ref.~\cite{Johannsen2015} for the deformation functions in the metric is based on an $M/r$ expansion (in units of $c=G=1$), where $M$ is the BH mass and $r$ the radial coordinate. This expansion becomes problematic at small $r$ for rapidly rotating BHs (e.g., near the inner edge of an accretion disk). This is exactly the regime where X-ray spectroscopy is most powerful~\cite{Dauser:2013xv,Abdikamalov:2019zfz} and typically used~\cite{Reynolds:2013qqa,Bambi:2018thh}. It is, therefore, essential to use a better metric to test the Kerr hypothesis, especially with X-ray reflection spectroscopy. In this paper, we present the application of a different metric to \textsc{relxill\_nk}, which overcomes these limitations. This metric has been proposed in Ref.~\cite{Konoplya2016}, and we will refer to it as the KRZ metric hereafter.}

\ch{The paper is organized as follows: in Sec.~\ref{s-review}, we review the KRZ metric and specify the deformation parameters, and review the \textsc{relxill\_nk} framework and the numerical methods involved. Sec.~\ref{sec:qual} presents a qualitative analysis of the effect of the deformation parameters on relevant quantities (e.g., the innermost stable circular orbit) and observables (e.g., the reflection spectrum). In Sec.~\ref{s-ana}, the new model is applied to X-ray observations of a supermassive BH, Ark~564. The findings of the Ark~564 data analysis are discussed and compared with estimates from other observational techniques in Sec.~\ref{s-discuss}. Also discussed in Sec.~\ref{s-discuss} are issues with systematic uncertainties and future works. Throughout the paper, we will express quantities in units of $c=G=1$. Except when mentioned explicitly, we will set the BH mass $M=1$.}

\section{Review\label{s-review}}
Non-Kerr metrics can be classified in two categories. The \textit{top-down} metrics are those which are obtained as a solution of an alternative theory of gravity, e.g., the Einstein-dilaton-Gauss-Bonnet BHs~\cite{Kanti1995,Kleihaus2015,Ayzenberg2014,Maselli2015,Kokkotas2017,Nampalliwar2018}, the Chern-Simons BHs~\cite{Yunes2009,Yagi2012,McNees2015,Delsate2018}, and the Kerr-Sen BHs~\cite{Sen1992,Hioki2008,Dastan2016,Uniyal2017}. The \textit{bottom-up} metrics on the other hand are obtained not from \ch{a specific theory} of gravity but by generalizing the Kerr metric~\cite{Vigeland2011,Johannsen2015,Konoplya2016,Papadopoulos:2018nvd,Carson:2020dez}. Each \ch{approach} has its advantages and disadvantages. Top-down metrics are difficult to obtain and \ch{might be known only numerically}, but testing for them amounts to testing an alternative theory of gravity. Bottom-up metrics may have pathologies in the spacetime, but they can be mapped to several top-down metrics and thus constraints on parameters of bottom-up metrics translates to constraints on several top-down metrics. \ch{Our focus in the present work is on a particularly attractive bottom-up case, the KRZ metric.}

\ch{\subsection{The KRZ metric \label{s-metric}}}
The KRZ metric~\cite{Konoplya2016} is based on \ch{a generic stationary and axisymmetric metric written in the Boyer-Lindquist like $(t,r,\theta,\phi)$ coordinates, where $t$ and $\phi$ are along the direction of a timelike and a spacelike Killing vector, respectively, $r$ and $\theta$ are orthogonal to each other and to $t$ and $\phi$, and at spatial infinity $(r,\theta,\phi)$ reduce to the standard spherical coordinates}. The metric functions are written in terms of continued fraction expansions in the polar and radial coordinates. 

\ch{The KRZ metric} has several advantages over other bottom-up metrics (See~\cite{Konoplya2016,Ni2016} for more details):
\begin{enumerate}
	\item Parametrizations based on expansion in $M/r$ suffer from the inherent weakness that close to the horizon, where effect of gravity are strongest, higher-order parameters become equally important, making it impossible to isolate the dominant terms in the expansion. The continued fraction expansion trick allows for quicker convergence with fewer parameters.
	\item A crucial feature of any bottom-up metric is its ability to map to several top-down metrics. The KRZ metric is generic enough to allow mapping to various top-down metrics with fewer parameters compared to other bottom-up metrics~\cite{Kokkotas2017,Nampalliwar2018,Konoplya:2020hyk}.
	\item Several bottom-up metrics are based on the Janis-Newman transformation, which is not guaranteed to work beyond general relativity~\cite{Hansen2013}. The KRZ metric uses a different approach to avoid this. 

\end{enumerate}
\ch{The line element of the KRZ metric reads~\cite{Konoplya2016,Ni2016}}
\bwide
\be\label{eq:metric}
ds^2 &=& - \frac{N^2 - W^2 \sin^2\theta}{K^2} \, dt^2 - 2 W r \sin^2\theta \, dt \, d\phi
+ K^2 r^2 \sin^2\theta \, d\phi^2 
+ \frac{\Sigma \, B^2}{N^2} \, dr^2 + \Sigma \, r^2 \, d\theta^2 \, ,
\ee
\ewide
where each function $N, B, W$ and $K$ depends on $r$ and $\theta$, and is written in a way that separates the terms that are constrained asymptotically (e.g., through parametrized post-Newtonian constraints) and those that are constrained near the BH horizon. For instance, 
\be
	B = 1 + \sum_{i=0}^{\infty} \left \{ b_{i0} (1-x) + \widetilde{B}_i (x) (1-x^2) \right \}y^i, 
\ee
where $x=1-r_0/r$ and $r_0$ is the BH horizon radius in the equatorial plane, $y=\cos\theta$, and the $b_{i0}$ terms are constrained to ensure $B$ has the correct asymptotic behavior. The $\widetilde{B}_i (x) $ terms encode all the strong-field deviations of the metric and are written as~\cite{Konoplya2016}
\be
 	\widetilde{B}_i (x) = \frac{b_{i1}}{1+\frac{b_{i2}x}{1+\frac{b_{i3}x}{1+\cdots}}}.
\ee

\ch{Since we are interested in parametric deformations away from the Kerr solution in the strong field regime, we will set all asymptotically constrained terms to their GR values and only retain the terms calculated close to the BH horizon. We will follow the choice made in Ref.~\cite{Ni2016} for the strong-field parameters to be retained. Assuming reflection symmetry across the equatorial plane and neglecting coefficients of higher orders, we get~\cite{Ni2016}}
\bwide
\be\label{eq:deffunc}
N^2 &=& \ch{\left(1 - \frac{r_0}{r}\right) \left (1 - \frac{\epsilon_0 r_0}{r} 
+ \left(k_{00} - \epsilon_0\right)\frac{r_0^2}{r^2} + \frac{\delta_1 r^3_0}{r^3}\right ) 
+ \left( \frac{a_{20} r^3_0}{r^3}
+ \frac{a_{21} r^4_0}{r^4} + \frac{k_{21} r^3_0}{r^3 \left ( 1+ \frac{k_{22}(1-\frac{r_0}{r}) }{1+k_{23}(1-\frac{r_0}{r})}\right)}\right) \cos^2\theta  \, ,} \\
B &=& 1 + \frac{\delta_4 r^2_0}{r^2} + \frac{\delta_5 r^2_0}{r^2} \cos^2\theta \, , \qquad
\Sigma = 1 + \frac{a_*^2}{r^2} \cos^2\theta \, , \\
W &=& \frac{1}{\Sigma} \left(\frac{w_{00} r^2_0}{r^2} + \frac{\delta_2 r^3_0}{r^3}
+ \frac{\delta_3 r^3_0}{r^3} \cos^2\theta \right) \, , \qquad \\
K^2 &=& 1 + \frac{a_* W}{r} + \frac{1}{\Sigma} \left( \frac{k_{00} r^2_0}{r^2} 
+ \left ( \frac{k_{20} r^2_0}{r^2} + \frac{k_{21} r^3_0}{r^3 \left ( 1+ \frac{k_{22}(1-\frac{r_0}{r})}{1+k_{23}(1-\frac{r_0}{r})}\right ) } \right ) \cos^2\theta \right)\, .
\ee
Here $a_*=J/M^2$ is the dimensionless spin parameter, 
\begin{align}
\begin{aligned}
r_0 = 1 + \sqrt{1 - a_*^2} \, , \qquad \epsilon_0 = \frac{2 - r_0}{r_0} \, ,  \qquad 
a_{20} = \frac{2 a_*^2}{r^3_0} \, , \qquad a_{21} = - \frac{a_*^4}{r^4_0} + \delta_6 \, ,\\
k_{00} = k_{22} = k_{23} = \frac{a_*^2}{r_0^2} \, , \qquad  k_{20} = 0 \, ,  \qquad
k_{21} = \frac{a_*^4}{r_0^4} - \frac{2 a_*^2}{r^3_0} - \delta_6 \, , \qquad w_{00} = \frac{2 a_*}{r^2_0} \, ,
\end{aligned}
\end{align}
\ewide
\ch{and $\{ \delta_i \}$ ($i = 1, 2, ... 6$) are six {\it deformation parameters}, which have been introduced to study deviations away from the Kerr metric. Note that the above expressions for the metric components are slightly different than those in Ref.~\cite{Konoplya2016}, since the expression given in Ref.~\cite{Konoplya2016} do not reduce to the Kerr metric in the limit of zero deformation. The expressions above do reproduce the Kerr solution exactly when all $\delta_i$ are set to zero.}
The physical interpretation of the deformation parameters can be summarized as follows (see Ref.~\cite{Konoplya2016} for more details):
\be
& \delta_1 & \rar \;\;\; \text{deformations of $g_{tt}$,}
\nonumber\\ 
& \delta_2 , \, \delta_3 & \rar \;\;\; \text{rotational deformations of the metric,}
\nonumber\\ 
& \delta_4 , \, \delta_5 & \rar \;\;\; \text{deformations of $g_{rr}$,}
\nonumber\\
& \delta_6 & \rar \;\;\; \text{deformations of the event horizon.}
\nonumber
\ee
With this choice, the mass-quadrupole moment is the same as in the Kerr metric, and deviations from the Kerr solution are only possible in the strong gravity region. 

\ch{Arbitrary deviations away from Kerr can create pathologies (e.g., closed time-like curves, Lorentz signature violation) in the spacetime. To avoid such scenarios, we impose restrictions on the above deformation parameters so that the following conditions are satisfied everywhere outside the horizon:
\begin{enumerate}
	\item The metric determinant is always negative.
	\item The metric coefficient $g_{\phi\phi}$ is greater than zero, and
	\item $N^2$ is non-vanishing.
\end{enumerate}
The explicit restrictions on each deformation parameter can be derived from these conditions using Eq.~\ref{eq:metric} and~\ref{eq:deffunc}. Assuming only one non-zero $\delta_i$ at a time, we get the following restrictions on each $\delta_i$:
\begin{gather}
	\delta_1 > \frac{4r_0 - 3r_0^2 - a^2}{r_0^2}, \\
	\delta_2, \delta_3 \left\{\begin{array}{l}
		> \\
		<\\
		\end{array} -\frac{4}{a^3}(1-\sqrt{1-a^2}) \quad \begin{array}{l}
		 \rm{if} \; a > 0\\ 
		 \rm{if} \; a < 0,
		\end{array} \right. \\
	\delta_4, \delta_5 > -1, \\
	\delta_6 < \frac{r_0^2}{4-a^2}.
\end{gather}
Note that currently \textsc{relxill\_nk} allows for variation in one deformation parameter at a time, so we will analyze the effect of each $\delta_i$ while fixing all other $\delta$s to zero. While mapping to BHs of alternative theories of gravity is unlikely to result in a single non-zero deformation parameter (e.g., see the mappings to static Einstein-dilaton-Gauss-Bonnet BHs in Ref.~\cite{Kokkotas2017} and slowly rotating Einstein-dilaton-Gauss-Bonnet BHs in Ref.~\cite{Konoplya2016}), we consider our analysis as a test of the validity of GR, in the spirit of null tests~\cite{Psaltis:2014mca,Ashby2018}.} 

\subsection{The Xspec model \label{s-relxillnk}}
	
	\begin{figure}[!htb]
		\includegraphics[width=0.48\textwidth]{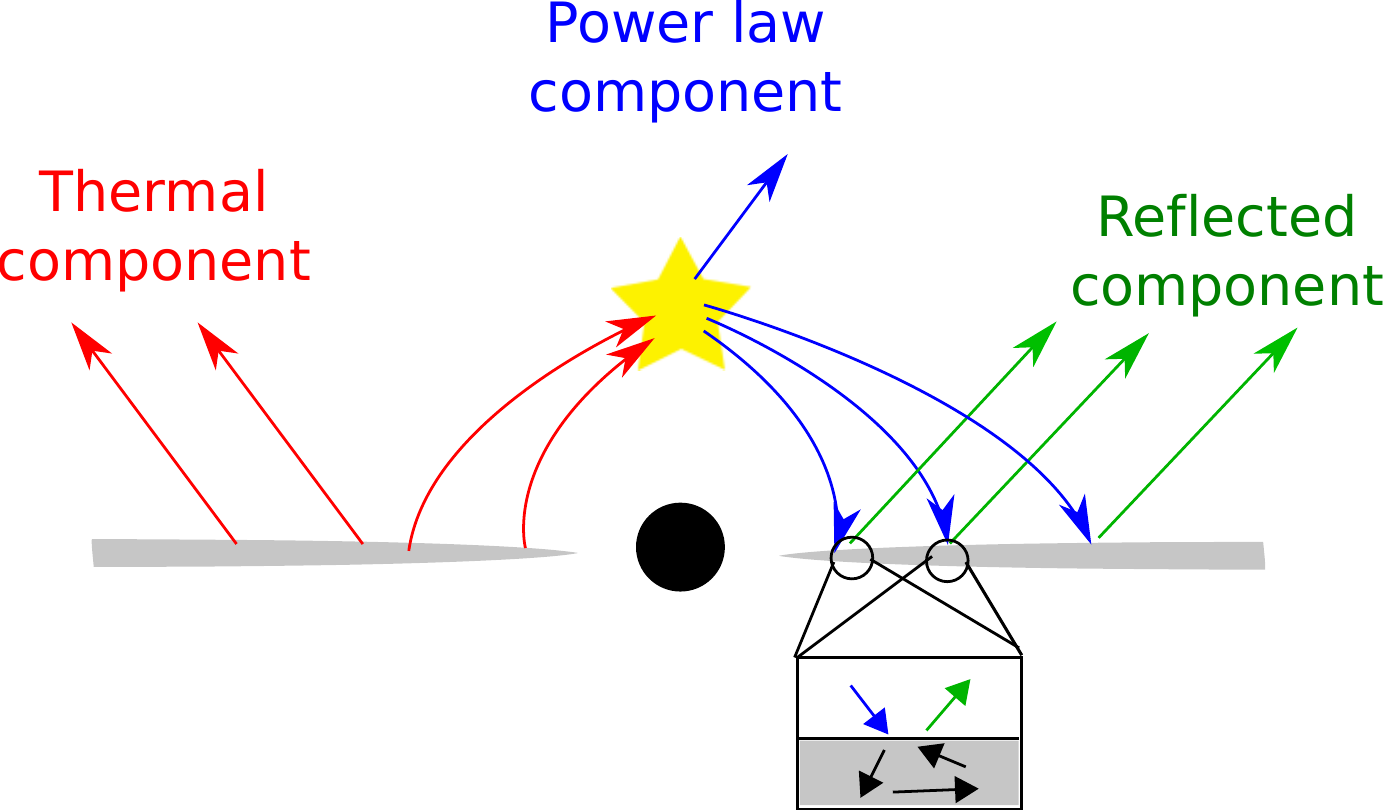}
		\caption{Schematic of the BH-accretion disk-corona system. The BH is colored black, the disk grey and the corona yellow. Various radiation components are labeled. The inset shows conversion of incident radiation in to reflected radiation.}
		\label{f-disk}
	\end{figure}
	
The standard astrophysical system in X-ray reflection spectroscopy is modeled with the BH-disk-corona model~\cite{Fabian:1989ej,Laor:1991nc,Bambi2017,Bambi:2018thh}. The BH is assumed to be surrounded by a geometrically thin and optically thick disk~\cite{Novikov1973},\footnote{Various other disk structures are possible. Studies of their effect on the reflection spectrum in the presence of non-Kerr metrics is an ongoing effort~\cite{Riaz:2019bkv,Abdikamalov:2020oci}.} with its inner edge at some radius $r_{\textrm{in}}$, often at the innermost stable circular orbit (ISCO), and the outer edge at some large radius $r_{\textrm{out}}$. In addition, the system possesses a ``corona'', which is thought to be a cloud of hotter (relative to the disk) gas or the base of an astrophysical jet. Fig.~\ref{f-disk} illustrates the system under discussion. The disk emits as a blackbody locally and as a multi-temperature blackbody when integrated radially (labeled the \textit{thermal component} in Fig.~\ref{f-disk}). Inverse Compton scattering of the thermal component by the corona produces X-rays (labeled the \textit{power-law component} in Fig.~\ref{f-disk}), some of which returns to the disk and is reflected (labeled the \textit{reflected component} in Fig.~\ref{f-disk}). Among these, the reflected component is most promising for studying the effects of strong gravity~\cite{Reynolds:2013qqa,Bambi2015}.
	
Modeling the reflection component requires some understanding of the various physical parameters of disk-corona model. \textsc{relxill\_nk} has several parameters to account for the different aspects of the system. These include the inner and outer edge of the disk, inclination of the disk relative to the observer, the disk's elemental constitution and their ionization, and the emissivity profile. The emissivity profile determines the reflection spectrum at the source, i.e., at the disk, and depends strongly on the coronal geometry. Since the latter is poorly understood, unless in specific cases like the lamp-post geometry, the emissivity profile is modeled by a power-law (intensity $\propto 1/r^q$) or a broken power-law (intensity $\propto 1/r^a$ for $r\le r_{\textrm{br}}$ and $\propto 1/r^b$ for $r> r_{\textrm{br}}$). \ch{Apart from the BH spin, the spacetime is determined by two parameters associated with deviations from Kerr: one parameter picks a specific  deformation parameter, e.g., KRZ $\delta_1$; the other parameter defines the value of the chosen deformation parameter.}\footnote{More details can be found on the public version webpage at Refs.~\cite{relxillnkweb,relxillnkweb2}.} \ch{This flexibility allows us to the use the same framework of \textsc{relxill\_nk} to analyze deformation parameters of various metrics. Note that \textsc{relxill\_nk} allows one deformation parameter at a time to vary, a limitation imposed by the FITS table file (see below) which grows exponentially with each additional deformation parameter.}

\subsection{Numerical method \label{s-trf}}
\ch{The numerical method has been presented in detail earlier~\cite{relxillnk,Abdikamalov:2019yrr}. Here we briefly review it. The \textsc{relxill\_nk} framework uses the transfer function formalism, introduced in ~\cite{Cunningham1975} and implemented in the \textsc{relxill} suite of models~\cite{Dauser:2013xv,Garcia:2013oma,Garcia:2013lxa} to optimize the calculation of the X-ray flux. In terms of the transfer function, the flux can be written as~\cite{relxillnk}}:
\bwide
\be\label{eq-Fobs}
F_o (\nu_o) 
= \int I_o(\nu_o, X, Y) d\tilde{\Omega} = \frac{1}{D^2} \int_{r_{\rm in}}^{r_{\rm out}} \int_0^1
\pi r_{ e} \frac{ g^2}{\sqrt{g^* (1 - g^*)}} f(g^*,r_{e},i)
I_{  e}(\nu_{ e},r_{ e},\vartheta_{ e}) \, dg^* \, dr_{ e} \, .
\ee
\ewide
Here $I_{ o}$ and $I_{ e}$ are the specific intensities of the radiation detected by the distant observer and the emitter respectively. These are related via the Liouville's theorem: $I_o = g^3 I_e$, where $g = \nu_o/\nu_e$ is the redshift factor, $\nu_o$ is the photon frequency as measured by the distant observer, and $\nu_e$ is the photon frequency in the rest frame of the emitter. $r_e$ is the emission radius in the disk and $\vartheta_{ e}$ is the photon's direction relative to the disk at the point of emission. $X$ and $Y$ are the Cartesian coordinates of the image of the disk in the plane of the distant observer, $D$ is the distance of the observer from the source, and $d\tilde{\Omega} = dX dY/D^2$ is the element of the solid angle subtended by the image of the disk in the observer's sky. The transfer function itself is defined as
\be\label{eq-trf}
f(g^*,r_e,i) = \frac{1}{\pi r_e} g 
\sqrt{g^* (1 - g^*)} \left| \frac{\partial \left(X,Y\right)}{\partial \left(g^*,r_e\right)} \right| \, .
\ee
Here $g^*$, the normalized redshift factor, is defined as
\be
g^* = \frac{g - g_{\rm min}}{g_{\rm max} - g_{\rm min}} \, ,
\ee
where $g_{\rm max}=g_{\rm max}(r_e,i)$ and $g_{\rm min}=g_{\rm min}(r_e,i)$ are, respectively, the maximum and the minimum values of the redshift factor $g$ at a constant $r_e$ and for a given viewing angle of the observer. The $r_e$-integral ranges from the inner to the outer edge of the disk, whereas the $g^*$-integral ranges from 0 to 1. \ch{Since there are two ways to go from $g^*=0$ to $g^*=1$ along a constant $r_e$ ring, there are two branches of the transfer function and the integral in Eq.~\ref{eq-Fobs} needs to be performed along both the branches. Furthermore, if the emission from the disk is not isotropic, then we also need to calculate $\vartheta_{ e}$ at each $r_e$ and $g^*$ along each branch, and Eq.~\ref{eq-Fobs} will be}
\bwide
\be\label{eq:flux}
	F_o (\nu_o) 
	= \frac{1}{D^2} \int_{r_{\rm in}}^{r_{\rm out}} \int_0^1 \pi r_{ e} \frac{ g^2}{\sqrt{g^* (1 - g^*)}} f^{(1)}(g^*,r_{e},i)
	I_{  e}(\nu_{ e},r_{ e},\vartheta_{ e}^{(1)}) \, dg^* \, dr_{ e} \, \nonumber \\
	+ \frac{1}{D^2} \int_{r_{\rm in}}^{r_{\rm out}} \int_0^1 \pi r_{ e} \frac{ g^2}{\sqrt{g^* (1 - g^*)}} f^{(2)}(g^*,r_{e},i)
	I_{  e}(\nu_{ e},r_{ e},\vartheta_{ e}^{(2)}) \, dg^* \, dr_{ e}.
\ee
\ewide
The transfer function separates the spacetime effects like the motion of gas and photons (encoded in the transfer function) from the local microphysics (encoded in the specific intensity at the emission point), and thus acts as an integration kernel for the calculation of flux. Such a separation enables quick computation of the reflection spectrum from a grid of transfer functions for any intensity profile, without the need to retrace photon trajectories, making analysis of X-ray reflection data with \textsc{relxill\_nk} possible in practice.

\ch{These transfer functions and emission angles are computed once and stored in a FITS (Flexible Image Transport System) table, and only the integral in Eq.~\ref{eq:flux} needs to be calculated during data analysis.} The table has a grid in three dimensions: spin $a_*$ \ch{(30 values)}, deformation parameter $\delta_i$ \ch{(30 values)}, and inclination angle $i$ \ch{(22 values)}. The grid points in the $a_*$ and $i$ dimension are non-uniform, mutually independent and follow the scheme in \textsc{relxill}. \ch{In particular, $a_*$ ranges from $-0.998$ to $0.9982$ and $i$ ranges from $3^{\circ}$ to $89^{\circ}$. The range in the $\delta_i$ dimension, in some cases, depends on the spin parameter: this is to ensure that the restrictions on $\delta_i$, given in Sec.~\ref{s-metric}, are followed. We use the following scheme for all $\delta_i$: The largest extent of $\delta_i$ at each spin is $-5$ to $5$, distributed uniformly between the two bounds. If the restriction given in Sec.~\ref{s-metric} is stronger than this bound, the range of that $\delta_i$ at that spin is set according to the restriction. Examples of this scheme can be seen in Fig.~\ref{fig:iscocon}, where the disallowed region is marked with stripes.}

\ch{At each grid point (namely a specific ($a_*$, $\delta_i$, $i$) on the grid described above), the accretion disk is discretized in 100 values of $r_e$, between $r_{\rm ISCO}$ and $1000M$, and 20 values of $g^*$ on each branch, between $10^{-3}$ and $1-10^{-3}$.\footnote{Since the Jacobian in the transfer function diverges at $g^* = 0$ and $1$, we offset the values by $10^{-3}$.} Photons are back-traced from the observer plane (placed at a large distance where the spacetime is effectively flat) to the accretion disk by solving the geodesic equations using a fourth-order Runge-Kutta scheme.} An adaptive algorithm fine-tunes the coordinates on the observer plane so that the photon when back-traced lands at the exact $r_{\textrm{e}}$. For each such ``central'' photon, the redshift and emission angle are calculated, and four photons, closely spaced in the observer plane, are launched to calculate the Jacobian and subsequently the transfer function. An interpolation routine then computes these quantities at the 20 values of $g^*$ and stores them in the FITS table.
\section{Qualitative analysis of deviation parameters\label{sec:qual}}

\begin{figure}[!htb]
		\centering
		\includegraphics[width=\columnwidth]{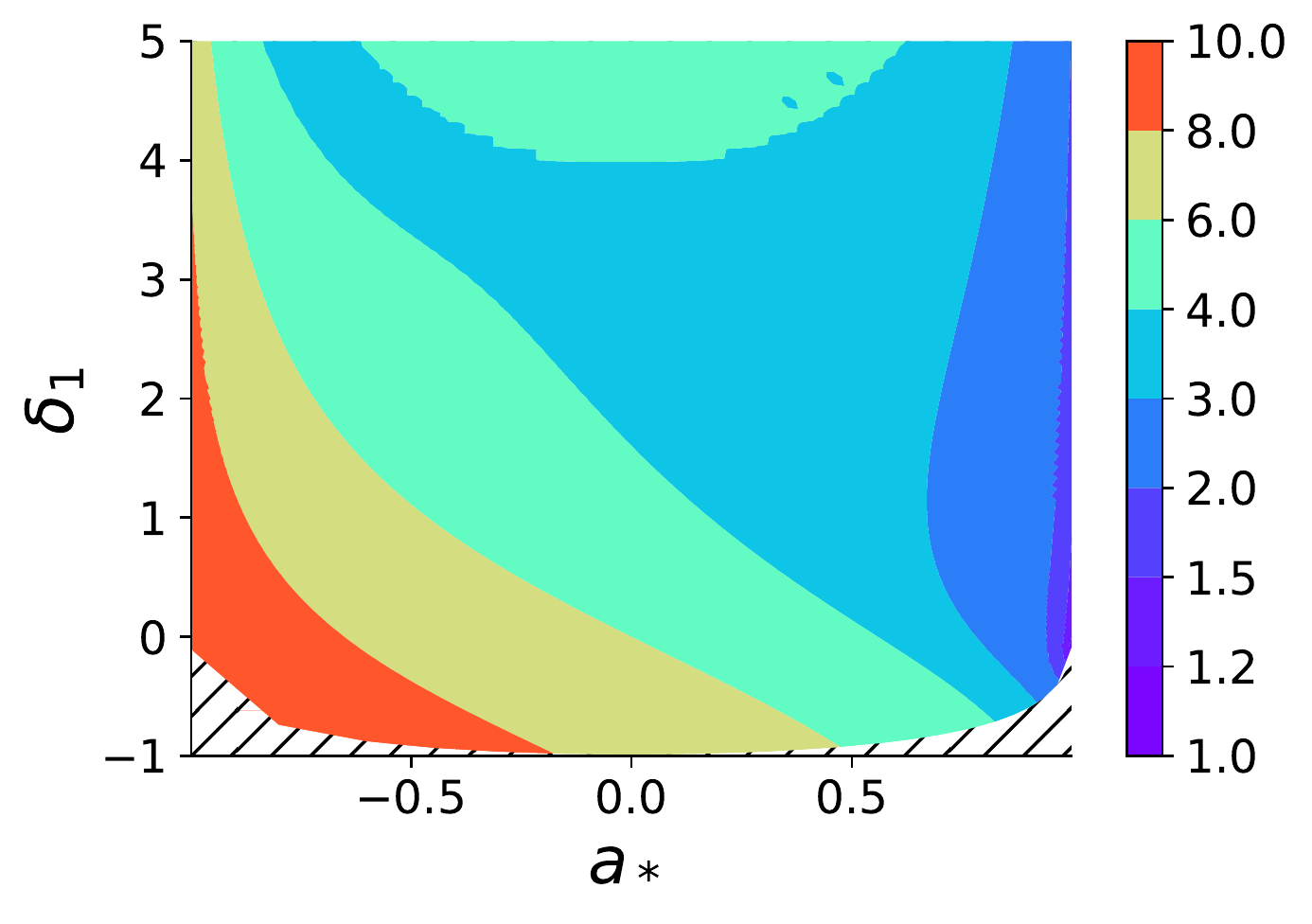}
		\includegraphics[width=\columnwidth]{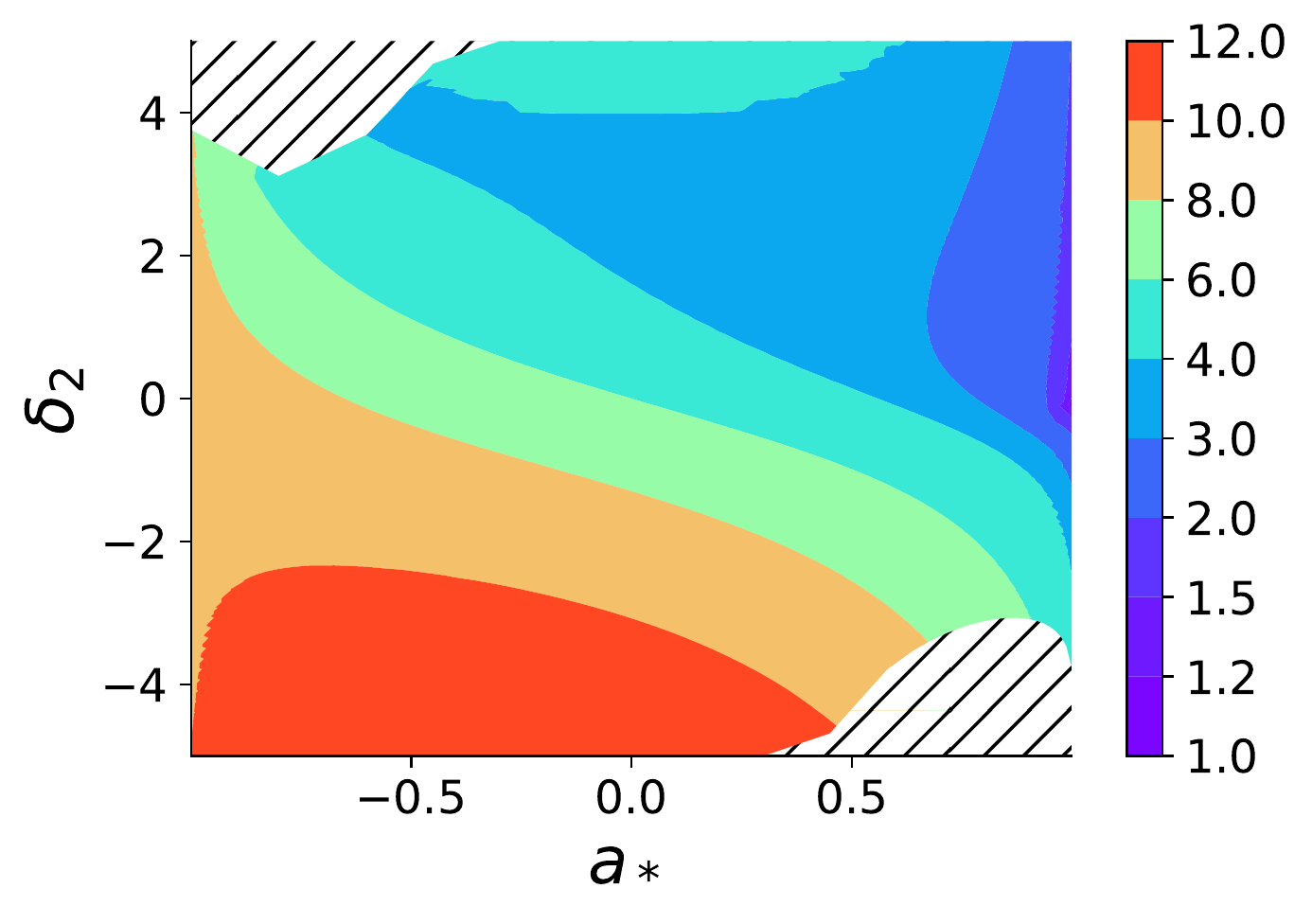}
		\caption{Impact of the deformation parameter on the ISCO radius. The horizontal axis shows the values of the BH spin, and the vertical axis shows the values of $\delta_1$ (top) and $\delta_2$(bottom), respectively. The colorbar gives the values of the ISCO radius. The hatched area is parameter space excluded from analysis (see Sec.~\ref{s-trf} and~\ref{s-metric}).}
		\label{fig:iscocon}
	\end{figure} 
\ch{We now look at the qualitative effect of the deformation parameters on some typical observables and quantities in X-ray reflection spectroscopy, to get some intuition how the observables are affected by the different kind of deformations, and anticipate when, if at all, they can be used to estimate the parameters.} 

\ch{One of the most important quantities is the ISCO. Since the emissivity profile has a radial fall-off, radiation from the innermost region has a significant influence on the X-ray spectra. The influence of the deformation parameters drops off radially as well, manifesting most prominently close to the horizon. Theoretically, the inner edge of the disk is assumed to be at ISCO within the Novikov-Thorne thin-disk models~\cite{Novikov1973}. Observationally too, disks in thermal state and accreting at $\sim5\%-30\%$ of the Eddington limit are expected to be bounded by the ISCO~\cite{McClintock:2006xd,Steiner:2010kd}. Thus, the ISCO serves as a proxy for determining constraints on deformation parameters with typical X-ray data. For axisymmetric metrics, the calculation of the ISCO radius involves only the $g_{tt}$, $g_{\phi\phi}$ and $g_{t\phi}$ components of the metric in the equatorial plane (see, for instance, Appendix A.3 in~\cite{relxillnk}). Thus, among the six deformation parameters under consideration here, only $\delta_1$ and $\delta_2$ influence the ISCO radius (see Eqs.~\ref{eq:metric}), while $\delta_3$, $\delta_4$, $\delta_5$, and $\delta_6$ have no effect (but they do affect other quantities like the redshift and the reflection spectrum, as we will see shortly). Fig.~\ref{fig:iscocon} shows how the ISCO radius changes with $\delta_1$ and $\delta_2$ in the top and bottom panels, respectively. Since the BH spin also has a strong effect on the ISCO, and in X-ray spectroscopy we typically measure both spin and deformation, each panel shows the variation in the ISCO radius on a spin-deformation parameter axes ($a_*-\delta_1$ in the top panel, $a_*-\delta_2$ in the bottom panel). In both cases, small ISCOs ($r_{\textrm{ISCO}}\lesssim 2M$) are confined to high spins, implying good estimates of $a_*$ are possible with X-ray data. For larger ISCO, the contours are spread over large ranges of $a_*$, making good spin estimates unlikely. For both $\delta_1$ and $\delta_2$, the ISCO contours stretch over a wide range, making precise estimation very difficult. The only exception is very small ISCOs ($r_{\textrm{ISCO}}\lesssim 1.5M$) for which both $\delta_1$ and $\delta_2$ are confined to a small range.}

\begin{figure*}[!htb]
		\centering
		\includegraphics[width=\columnwidth]{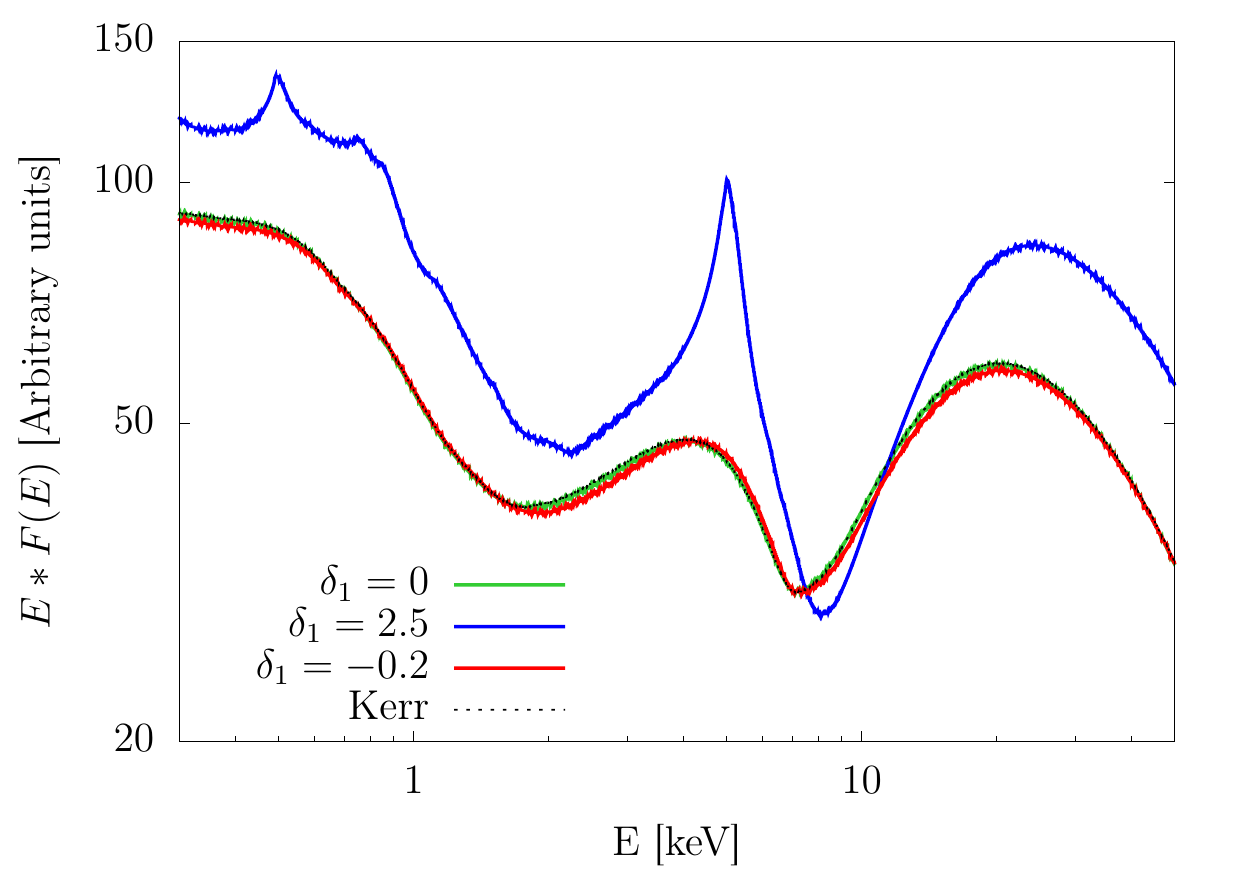}
		\includegraphics[width=\columnwidth]{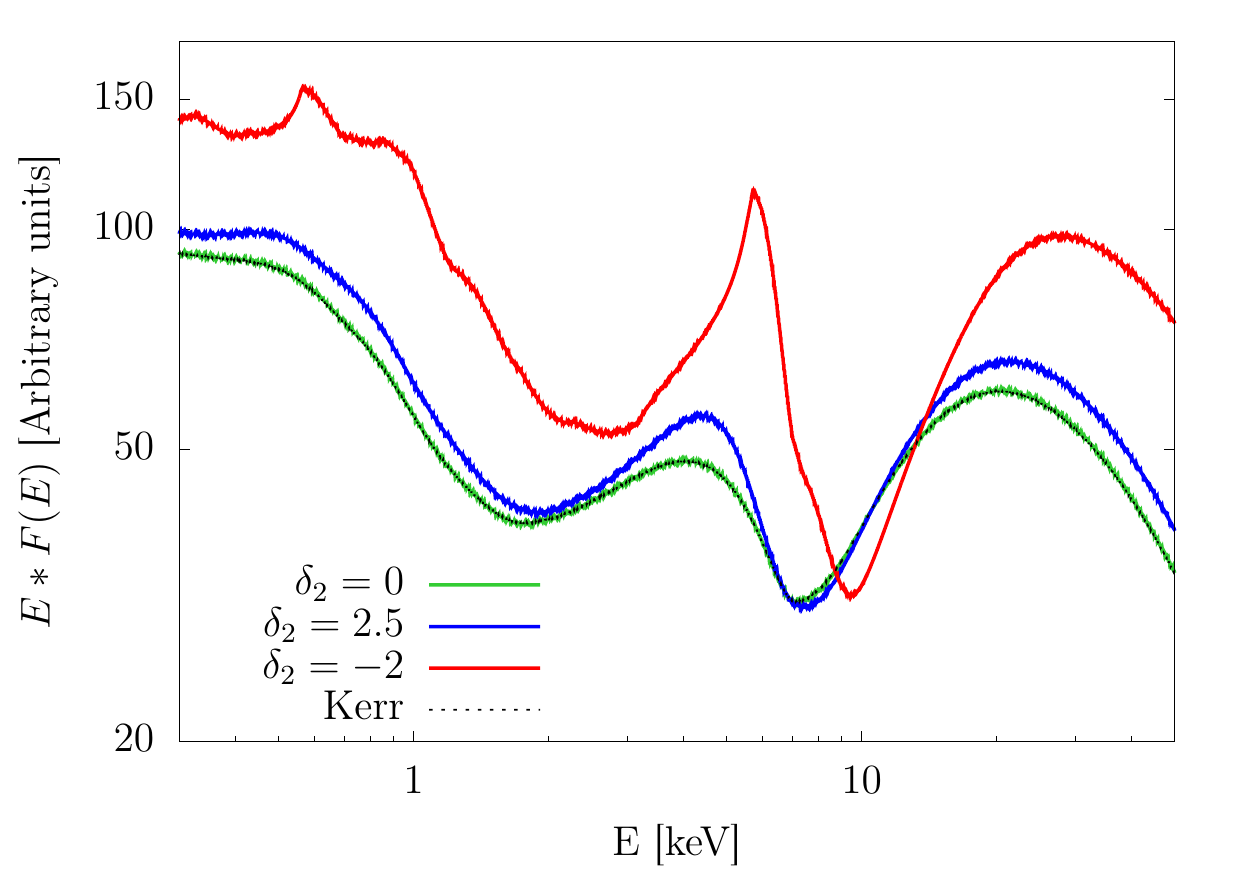}
		\includegraphics[width=\columnwidth]{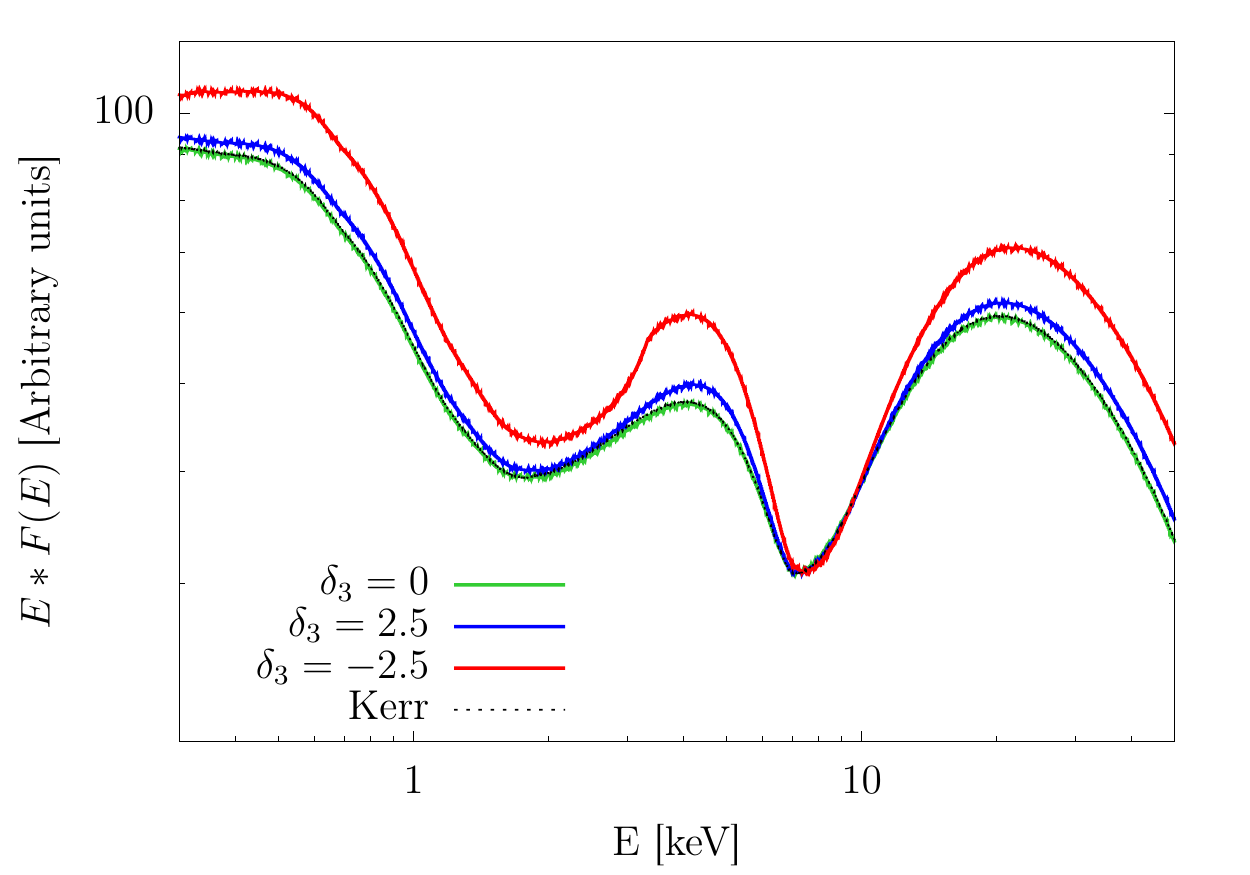}
		\includegraphics[width=\columnwidth]{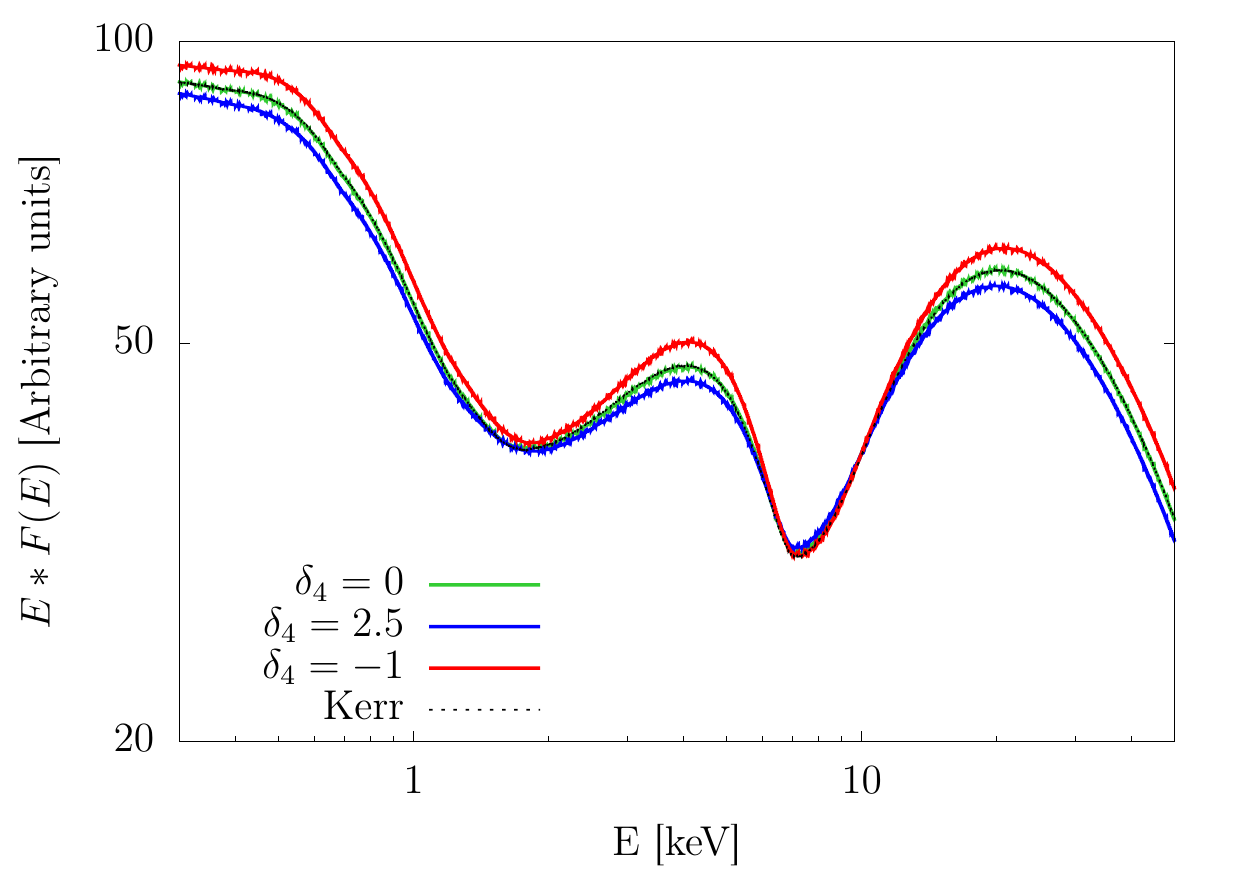}
		\includegraphics[width=\columnwidth]{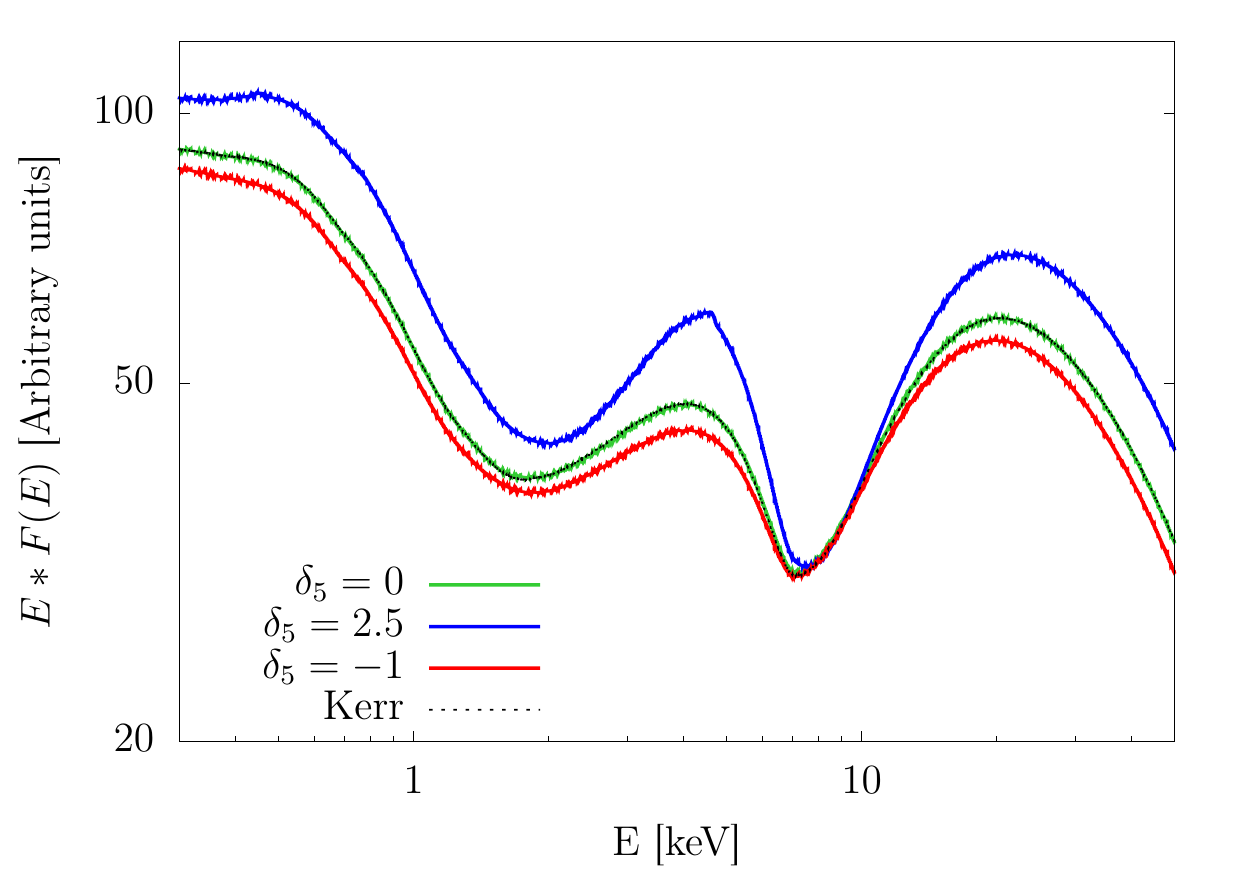}
		\includegraphics[width=\columnwidth]{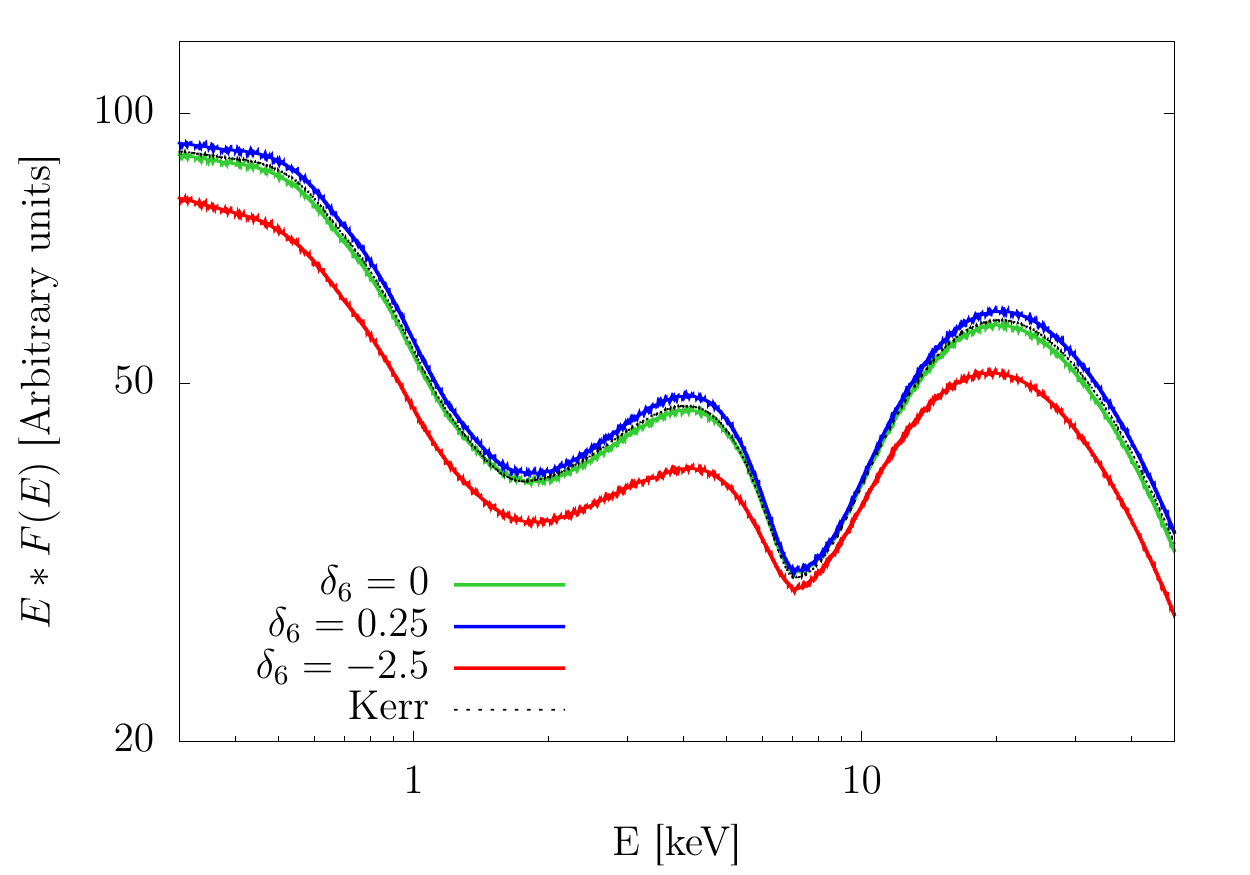}
		\caption{Impact of $\delta_i$ on the reflection spectrum. Also plotted is the spectrum in the Kerr case, for comparison with $\delta_i=0$ in each case. The fixed parameters are: spin $=0.99$, inclination $=30^\circ$, ionization $\log\xi=3.1$, iron abundance $A_{\mathrm{Fe}}=5$, photon index $\Gamma=2$, cut-off energy $E_{\mathrm{cut}}=300$ keV, and emissivity index $q=6$. The disk extends from $r_{\textrm{in}}=r_{\textrm{ISCO}}$ to $r_{\textrm{out}}=400M$.}
		\label{fig:spectra}
\end{figure*}

\ch{We now look at the primary observable in X-ray reflection spectrocopy and the output of the \textsc{relxill\_nk} model. Fig.~\ref{fig:spectra} shows the reflection component of the flux from a typical BH-disk system, for various $\delta_i$s. Each panel shows one specific $\delta_i$, and within each panel spectra for three different values of the $\delta_i$ parameter are shown. Remaining model parameters are kept constant in all panels (and listed in the figure caption). In particular, $a_*$ is set to $0.99$ and $q$ to $6$, since X-ray reflection spectroscopy is most useful for performing tests of gravity in systems with very high BH spins and steep emissivity profiles~\cite{Fabian:2002gj,Dauser:2013xv}. In each panel, we plot the spectra for zero deformation ($\delta_i=0$) and positive/negative deformation ($\delta_i \lessgtr 0$). The positive/negative deformation value is typically set to $\pm 2.5$, except when the range of $\delta_i$ restricts it to a smaller deformation (see, e.g., Fig.~\ref{fig:iscocon} and Sec.~\ref{s-trf}). Several features can be seen in the plots. The spectrum is very sensitive to positive $\delta_1$ while negative values of $\delta_1$ are severely restricted and the spectrum does not change much for the most negative $\delta_1$ allowed. In the $\delta_2$ case, the spectrum is very sensitive to negative values of $\delta_2$, while positive values have a moderate effect. $\delta_3$
also has a moderate-to-strong effect on the spectrum for negative values, whereas positive $\delta_3$ affects the spectrum rather weakly. $\delta_4$ has negligible effect for either positive or negative values, but $\delta_5$ has a strong effect for positive values. Finally, $\delta_6$ has a strong effect on the spectrum for negative values, whereas positive values do not affect the spectrum much. In each panel, we also plot the spectra from \textsc{relline}~\cite{Dauser:2010ne}, which includes only the Kerr metric, to facilitate comparison of the zero deformation case with the Kerr case. For all $\delta_i$s, the zero deformation case matches very well with the Kerr case.}

\ch{The ability of X-ray spectroscopy, and in fact any technique, to constrain parameters depends not only on the effect of the parameter on the relevant observable, but also on the degeneracy among the parameters in terms of their effect on the observable. In particular, above we looked at the effect of $\delta_i$ on the reflection spectra at fixed BH spin. But it is well known that spin and deformation parameter affect the spectra in a somewhat similar way, resulting in a degeneracy. For statistically robust parameter estimation, it is critical to ensure all degeneracies are accounted for. We will see instances of this in the next sections where we estimate $\delta_i$ from astrophysical data.}

\section{Spectral analysis of Ark~564 data\label{s-ana}}

In this section, we use our newly developed model to analyze an X-ray observation of the supermassive BH in Ark~564~\cite{Walton2012}. 

\subsection{Review \label{ss:ana-rev}}
Ark~564 is classified as a narrow line Seyfert 1 galaxy at redshift $z = 0.0247$. Since first observations with \textsl{XMM-Newton} in 2000/2001~\cite{Vignali:2003hn}, it has been studied by several authors since it appears as a very bright source in the X-ray band~\cite{Papadakis:2006ms,McHardy:2007qg,Kara:2017jdb,Barua:2020hhd}. It is a good first candidate for tests of general relativity with a new metric for the following reasons. Firstly, previous studies have shown that the inner edge of the disk may be very close to the central object, which maximizes the signatures of the strong gravity region~\cite{Walton2012}. Secondly, the source has a simple spectrum. There is no obvious intrinsic absorption to complicate the determination of the reflected emission. The same dataset was analyzed by some of us in Ref.~\cite{Tripathi2018a} to constrain the Johannsen parameters $\alpha_{13}$ and $\alpha_{22}$, and in Ref.~\cite{Tripathi:2019fms} to constrain the Johannsen parameter $\epsilon_{3}$. Therein, this source and this dataset was found to be easy to analyze and provided good constraints on the deformation parameters. There are other X-ray sources with the potential to provide stronger constraints than this data (e.g. GRS 1915+105~\cite{Zhang:2019ldz}, MCG-06-30-15~\cite{Tripathi:2018lhx}) and their analysis with the KRZ metric is underway.

\subsection{Observations and data reduction \label{s-red}}

\textsl{Suzaku} observed Ark 564 on 26-28 June 2007 (Obs. ID 702117010) for about $80$~ks. For low energies ($< 10$~keV), \textsl{Suzaku} has four co-aligned telescopes which are used to collect photons onto its CCD detectors X-ray Imaging Spectrometer (XIS). XIS is comprised of four detectors; XIS0, XIS2, and XIS3 are front-illuminated and XIS1 is back-illuminated. We only used data from the front-illuminated chips because XIS1 has a lower effective area at 6~keV and a higher background at higher energies. XIS2 data were not used in our analysis because of the anomaly after 9 November 2006.

We have used HEASOFT version 6.24 and CALDB version 20180312 for the data reduction. AEPIPELINE script of the HEASOFT package has been used for reprocessing and screening of the raw data. The ftool XSELECT and ds9 were used to extract the XIS source and background spectrum from a 3.5 arc minute radius. The background region was selected as far as possible from the source so as to avoid contamination. The RMF file was generated using XISRMFGEN and the ARF file which corresponds to effective area of the telescope was generated using XISSIMARFGEN. At last, the final source spectrum, background spectrum and response file were generated by combining the data from XIS0 and XIS3 using ADDASCASPEC. The data was then grouped using GRPPHA to get minimum 50 counts per bin so as to use $\chi^2$ statistics in our spectral analysis. We have also excluded the energy range between 1.7-2.5 because of calibration issues.

\subsection{Modelling and results \label{s-results}}
\begin{figure*}[!htb]
    \centering
    \includegraphics[width=0.9\textwidth]{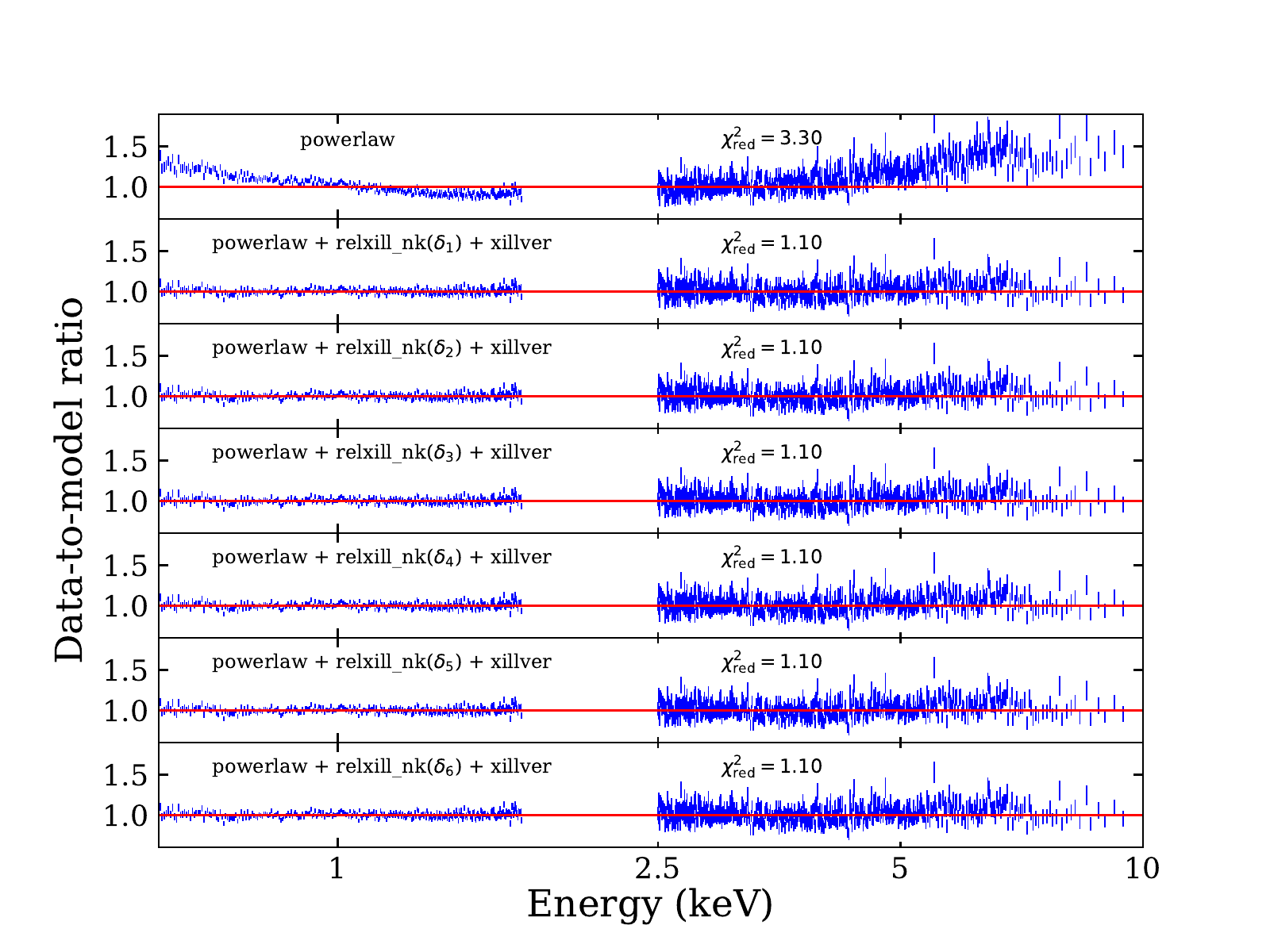}
    \caption{Data to best-fit model ratios for model 1 (top panel) and 2 for each $\delta_i$ (rest of the panels), as defined in Sec.~\ref{s-red}. The corresponding parameter estimates are listed in Tab.~\ref{tab:params}. See the text for more details.}
    \label{fig:fig1}
\end{figure*}

\ch{We fit the data with Xspec v12.10.0c. The data has been analyzed before, with models that incorporate the Kerr metric~\cite{Kara:2017jdb} and with those incorporate non-Kerr metrics~\cite{Tripathi2018a,Tripathi:2019fms}. In particular, we will use the findings of Ref.~\cite{Tripathi2018a}, where a comprehensive analysis has been done of the model combinations, and the optimal combination found had a relativistic and a non-relativistic reflection component. Of course, there is no \textit{a priori} reason to assume that the models used to fit some data must be the same for all non-Kerr parameters, so we follow the standard approach by first fitting the power-law component (see Sec.~\ref{s-relxillnk} and Fig.~\ref{f-disk}), followed by the reflection components.}

\begin{center}
Model 1 \\
\textsc{tbabs*zpowerlw}
\end{center}

\textsc{tbabs} describes the galactic absorption~\cite{Wilms2000} and we fix
the galactic column density to $N_H = 6.74 \cdot10^{20} \textrm{cm}^{-2}$~\cite{galabsweb}. \textsc{zpowerlw} describes a redshifted photon powerlaw spectrum. \ch{The best-fit values are reported in the second column in Tab.~\ref{tab:params}, and the data-to-model ratio plots are presented in the top panel in Fig.~\ref{fig:fig1}. An excess of photons at low energies and a broad iron line around 6.4 keV can be seen in the panel, suggesting the presence of a reflected component.} Invoking this paradigm, we proceed to fit the spectrum using the reflection models \textsc{relxill\_nk} and \textsc{xillver}.

\begin{center}
Model 2\\\textsc{tbabs*(zpowerlw + relxill\_nk + xillver)}
\end{center}

\ch{The power-law component is still modeled with \textsc{zpowerlw} as before, while the reflection is modeled with \textsc{relxill\_nk} and \textsc{xillver}. For the fit, the disk is assumed to extend from the ISCO to $400M$. The emissivity profile is assumed to be a simple power law, whose index $q$ is a free parameter. 
The photon index in \textsc{relxill\_nk} is tied to \textsc{zpowerlw} and the cut-off energy frozen to the default value ($300$ keV). The ionization and iron abundance of the disk, along with the BH spin and deformation parameter, are kept free. 
The parameters of \textsc{xillver} are mostly fixed or tied (ionization fixed to zero, iron abundance, cut-off energy, inclination tied to \textsc{relxill\_nk}, photon index tied to \textsc{zpowerlw}), the only free parameter being the norm. The fit is performed six times, once for each $\delta_i$.
}

\ch{During the fit, inclination had to be handled with care, for the following reason: previous constraints on inclination with this (or combined with this) observation of Ark~564 have not been very good, either unbounded from below (see Tab.~I of Ref.~\cite{Tripathi2018a} and Tab.~1 of Ref.~\cite{Kara:2017jdb}) or very precise but inconsistent (see Tab.~II of Ref.~\cite{Tripathi2018a}, Tab.~II of Ref.~\cite{Tripathi:2019fms} and Tab.~1 of Ref.~\cite{Kara:2017jdb}). Keeping inclination free in our analysis, we run into similar issues, i.e., either the parameter is largely unbounded or stuck at the smallest allowed value ($\sim3^{\circ}$) with little uncertainty. Instead, we bound it to lie around $30^{\circ}$ with a $10\%$ uncertainty (i.e., $27^{\circ}<i<33^{\circ}$), which encompasses to a large extent estimates obtained with a different observation of Ark~564~\cite{Kara:2017jdb,Barua:2020hhd} made with \textsl{NuSTAR}, and with the \textsl{Suzaku} observation and different deformation parameters~\cite{Tripathi2018a,Tripathi:2019fms}.} 

\begin{table*}[t]
\renewcommand{\arraystretch}{1.4}
 \begin{center}
  \begin{tabular}{m{2.7cm} >{\centering\arraybackslash}m{2cm} >{\centering\arraybackslash}m{2cm} >{\centering\arraybackslash}m{2cm} >{\centering\arraybackslash}m{2cm} >{\centering\arraybackslash}m{2cm} >{\centering\arraybackslash}m{2cm} >{\centering\arraybackslash}m{2cm}}
   \hline
   \textbf{Model} & \textbf{1} & \textbf{2}$(\delta_1)$ & \textbf{2}$(\delta_2)$ & \textbf{2}$(\delta_3)$ & \textbf{2}$(\delta_4)$& \textbf{2}$(\delta_5)$ & \textbf{2}$(\delta_6)$ \\
   \hline \hline
   \textsc{tbabs} \\
   $N_H/10^{20} cm^{-2}$ & $6.74^*$ & $6.74^*$ & $6.74^*$ & $6.74^*$ & $6.74^*$ & $6.74^*$ & $6.74^*$\\
   \hline
   \textsc{zpowerlaw} \\
   $\Gamma$ & $2.78^{+\epsilon}_{-0.01}$ & $2.86^{+0.05}_{-0.04}$ & $2.86^{+0.06}_{-0.04}$ & $2.86^{+0.06}_{-0.04}$ & $2.85^{+0.01}_{-0.03}$ & $2.87^{+0.04}_{-0.03}$ & $2.87^{+0.05}_{-0.02}$ \\
   $N$ & $0.68$ & $0.39^{+0.03}_{-0.04}$ & $0.39^{+0.02}_{-0.04}$ & $0.39^{+0.03}_{-0.06}$ & $0.39^{+0.02}_{-0.03}$ & $0.39^{+0.02}_{-0.03}$ & $0.39^{+0.02}_{-0.04}$\\
   \hline
   \textsc{relxill\_nk} \\
   $q$ & - & $8.20^{+\mathrm{P}}_{-0.05}$ & $8.66^{+\mathrm{P}}_{-0.05}$ & $8.57^{+0.79}_{-0.49}$ & $8.52^{+0.67}_{-0.36}$ & $8.83^{+0.44}_{-0.63}$ & $8.39^{+0.48}_{-0.36}$ \\
   $a_*$ & - & $0.998^{+\mathrm{P}}_{-0.022}$ & $0.998^{+\mathrm{P}}_{-0.010}$ & $0.998^{+\mathrm{P}}_{-0.012}$ & $0.986^{+0.009}_{-0.022}$ & $0.986^{+0.006}_{-0.003}$ & $0.986^{+0.005}_{-0.013}$\\
   $i$ [deg] & - & $27^{+2}_{-\mathrm{P}}$ & $27^{+3}_{-\mathrm{P}}$ & $27^{+\mathrm{P}}_{-\mathrm{P}}$ & $27^{+4}_{-\mathrm{P}}$ & $29^{+\mathrm{P}}_{-\mathrm{P}}$ & $27^{+\mathrm{P}}_{-\mathrm{P}}$ \\
   $\log\xi$ & - & $3.00^{+0.12}_{-0.53}$ & $3.00^{+0.07}_{-0.18}$ & $3.00^{+0.09}_{-0.24}$ & $3.00^{+0.04}_{-0.16}$ & $3.00^{+0.03}_{-0.16}$ & $2.98^{+0.08}_{-0.11}$ \\
   $A_{\textrm{Fe}}$ & - & $0.83^{+0.13}_{-0.18}$ & $0.84^{+0.14}_{-0.09}$ & $0.85^{+0.14}_{-0.18}$ & $0.85^{+0.07}_{-0.15}$ & $0.81^{+0.09}_{-0.09}$ & $0.88^{+0.09}_{-0.09}$ \\
   $\delta$-type & - 		& $\delta_1$ & $\delta_2$ & $\delta_3$ & $\delta_4$ & $\delta_5$ & $\delta_6$\\
   $\delta$-value & - & $0.197^{+0.087}_{-0.464}$ & $-0.254^{+0.469}_{-0.117}$ & $-0.414^{+1.452}_{-0.035}$ & $2.928^{+\mathrm{P}}_{-3.475}$ & $-0.896^{+0.818}_{-\mathrm{P}}$ & $-0.610^{+0.510}_{-4.262}$\\
   $N/10^{-2}$ 	& - 	& $3.56^{+3.61}_{-0.52}$ & $3.51^{+0.84}_{-0.28}$ & $3.51^{+3.06}_{-0.44}$ & $3.28^{+0.50}_{-0.33}$ & $3.91^{+0.31}_{-0.28}$ & $3.85^{+0.31}_{-0.28}$ \\
   \hline
   \textsc{xillver} \\
   $N/10^{-3}$ & - & $5.20^{+1.32}_{-1.64}$ & $5.26^{+1.80}_{-1.59}$ & $5.25^{+1.83}_{-1.66}$ & $5.28^{+1.59}_{-1.74}$ & $5.32^{+1.62}_{-1.80}$ & $5.16^{+1.62}_{-1.80}$ \\
   \hline
   \hline
   $\chi^2/dof$ & $3.30$ & $1.10$ & $1.10$ & $1.10$ & $1.10$ & $1.10$ & $1.10$ \\
   \hline
   \hline
  \end{tabular}
    \caption{\ch{Summary of the best-fit values for the spectral models 1 and 2, the latter for each $\delta_i$ case. Uncertainties are reported at 90\% confidence level. All values are reported up to two significant digits after the decimal, except $a_*$ and $\delta_i$ which are reported to three significant digits after the decimal. $\mathrm{P}$ indicates the relevant parameter being unbounded up to its explored range and $\epsilon$ indicates uncertainty too small up to the precision reported here. See the text for more details.}}
    \label{tab:params}
 \end{center}
\end{table*}

\ch{The data-to-model ratios are plotted for each iteration of Model 2, i.e., for each $\delta_i$, in Fig.~\ref{fig:fig1}. For each best-fit model, parameter values are listed in Tab.~\ref{tab:params}, with associated uncertainty at $90\%$ confidence level (CL). We find that, in all cases, the fit is good with a $\chi_{\textrm{red}}^2$ close to 1 and no major unmodeled features in the ratio plots. We find a high value for the photon index $q$, that is, most of the radiation seems to come from very inner part of the accretion disk, consistent with previous analyses~\cite{Tripathi2018a,Tripathi:2019fms}. The photon index is slightly higher than Ref.~\cite{Kara:2017jdb} (both being higher than in Ref.~\cite{Barua:2020hhd}) but is consistent for each $\delta_i$ case. Ionization is consistent amongst the Model 2 iterations and close to previous analyses (lower than Ref.~\cite{Tripathi:2019fms}, higher than Ref.~\cite{Tripathi2018a}),   
The spin parameter $a_*$ is always very close to 1. This is consistent with the previous analysis results of Ark~564 and suggests that it is a high spin BH. Iron abundance is close to solar and matches well with Ref.~\cite{Tripathi2018a} (see also Ref.~\cite{Kara:2017jdb}). The inclination is, in most cases, pegged to the smallest allowed value ($27^{\circ}$) and, in some cases, unbounded within its restricted range.}

\section{Discussion\label{s-discuss}}
\begin{figure*}[!htb]
    \centering
    \includegraphics[width=\columnwidth]{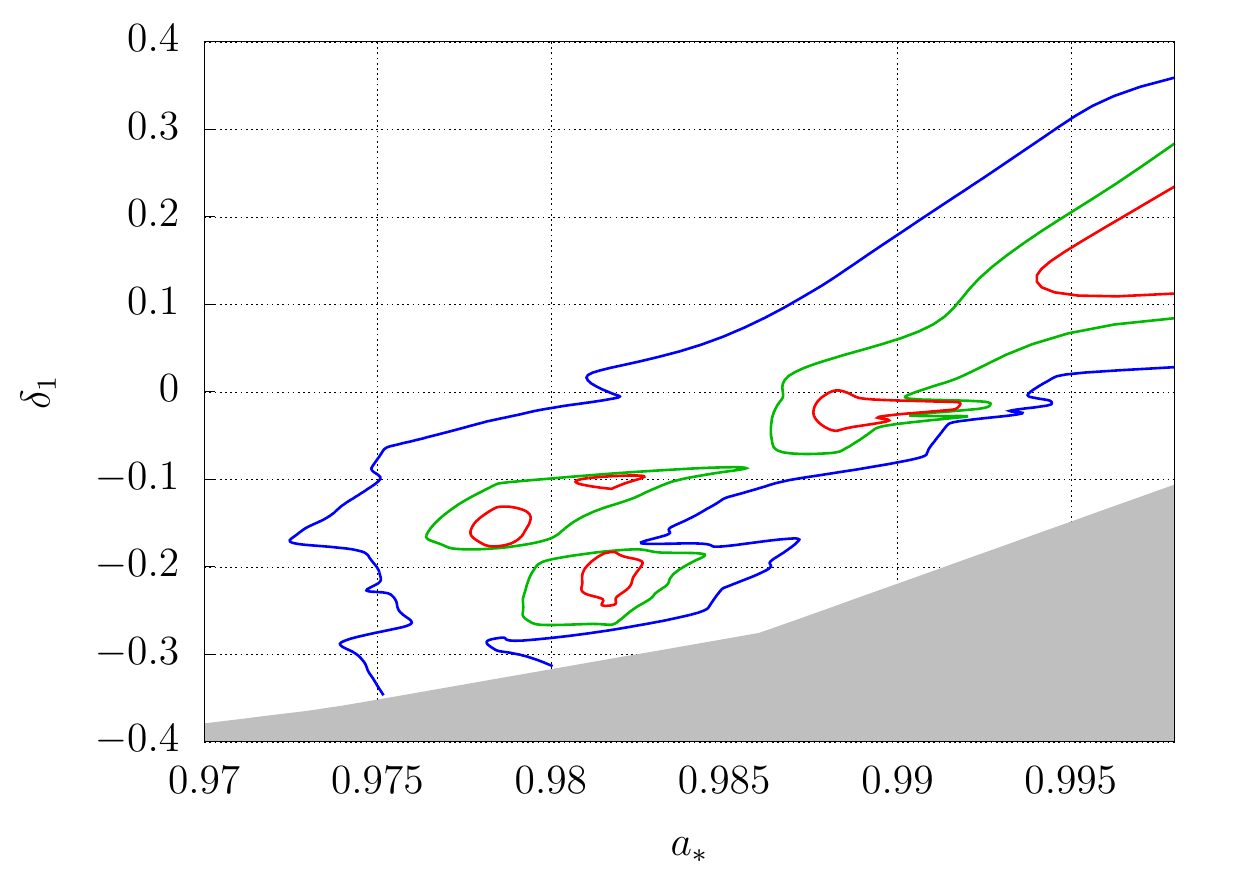}
     \includegraphics[width=\columnwidth]{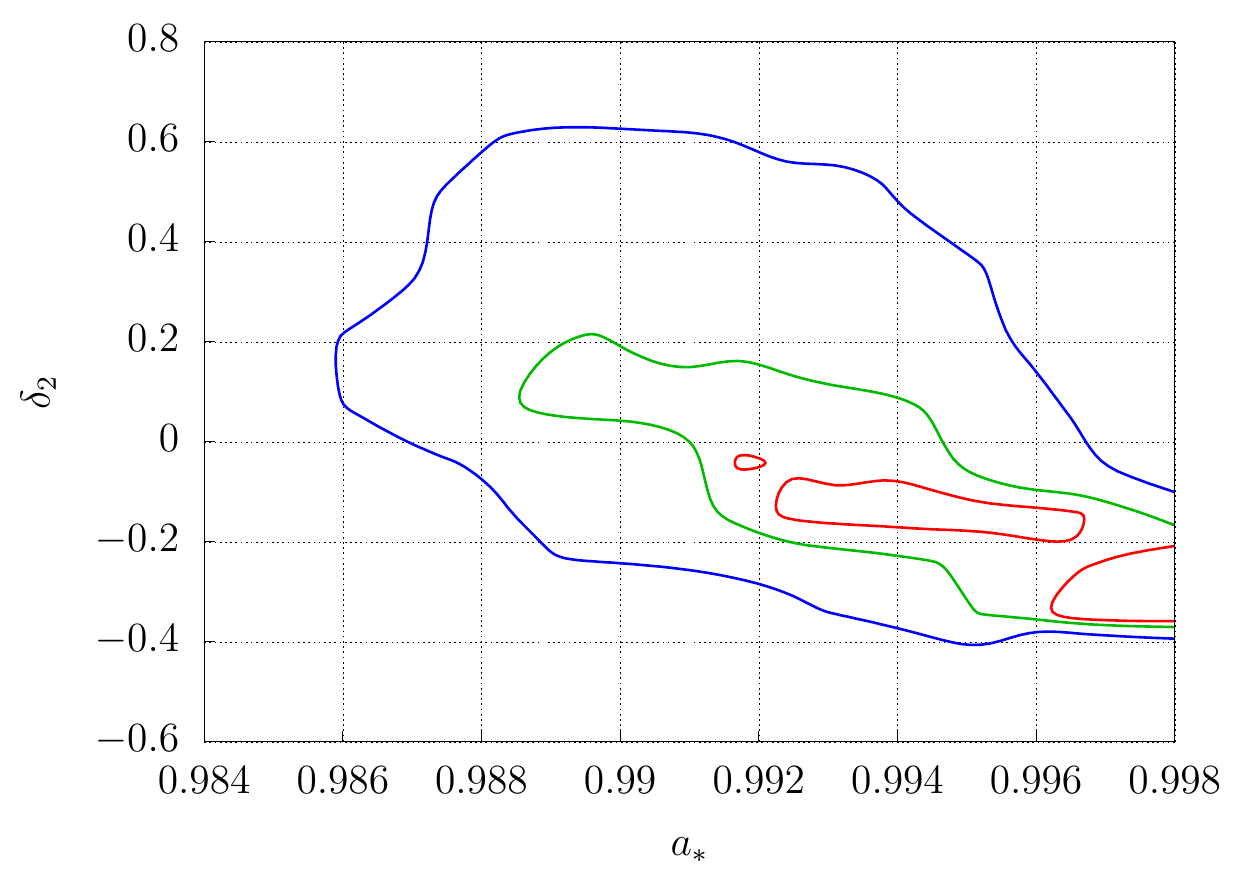}
     \includegraphics[width=\columnwidth]{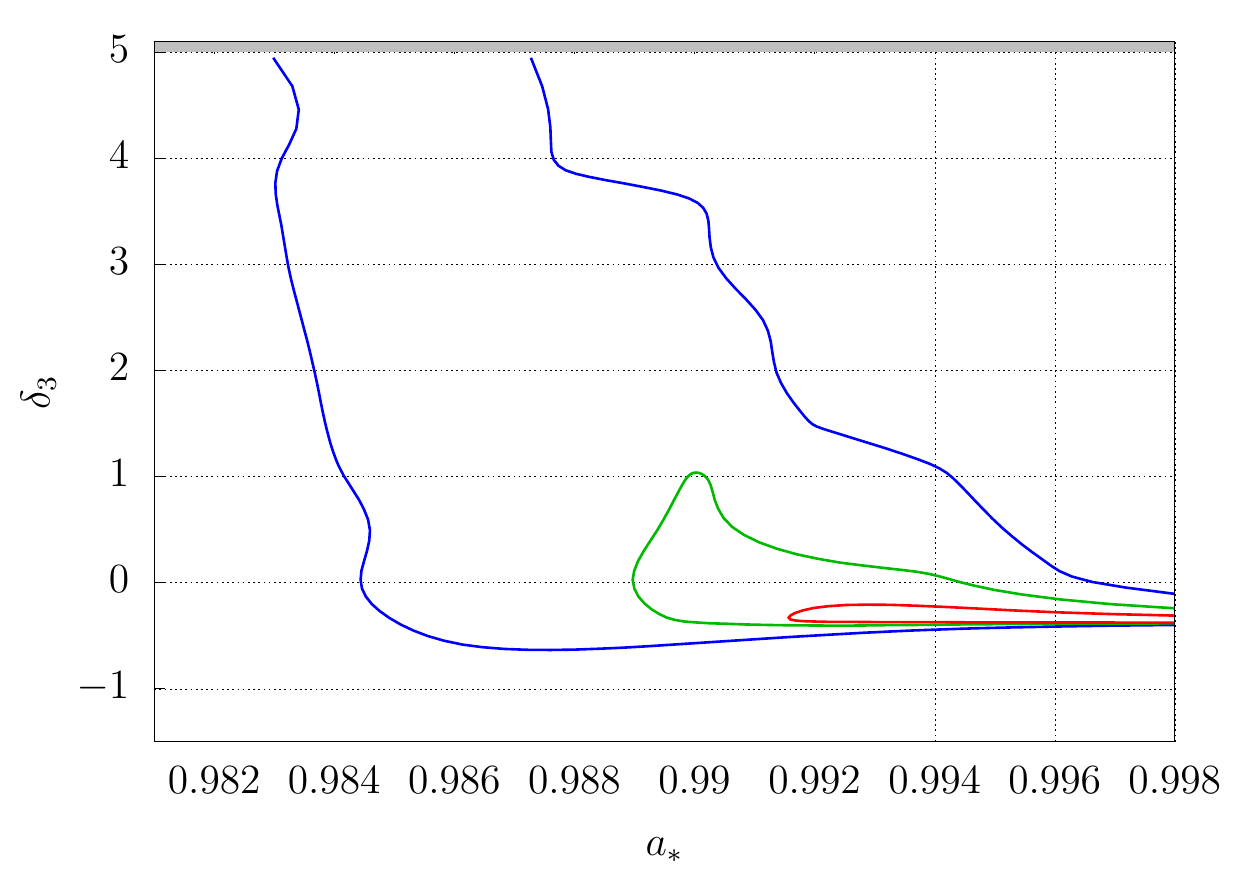}
     \includegraphics[width=\columnwidth]{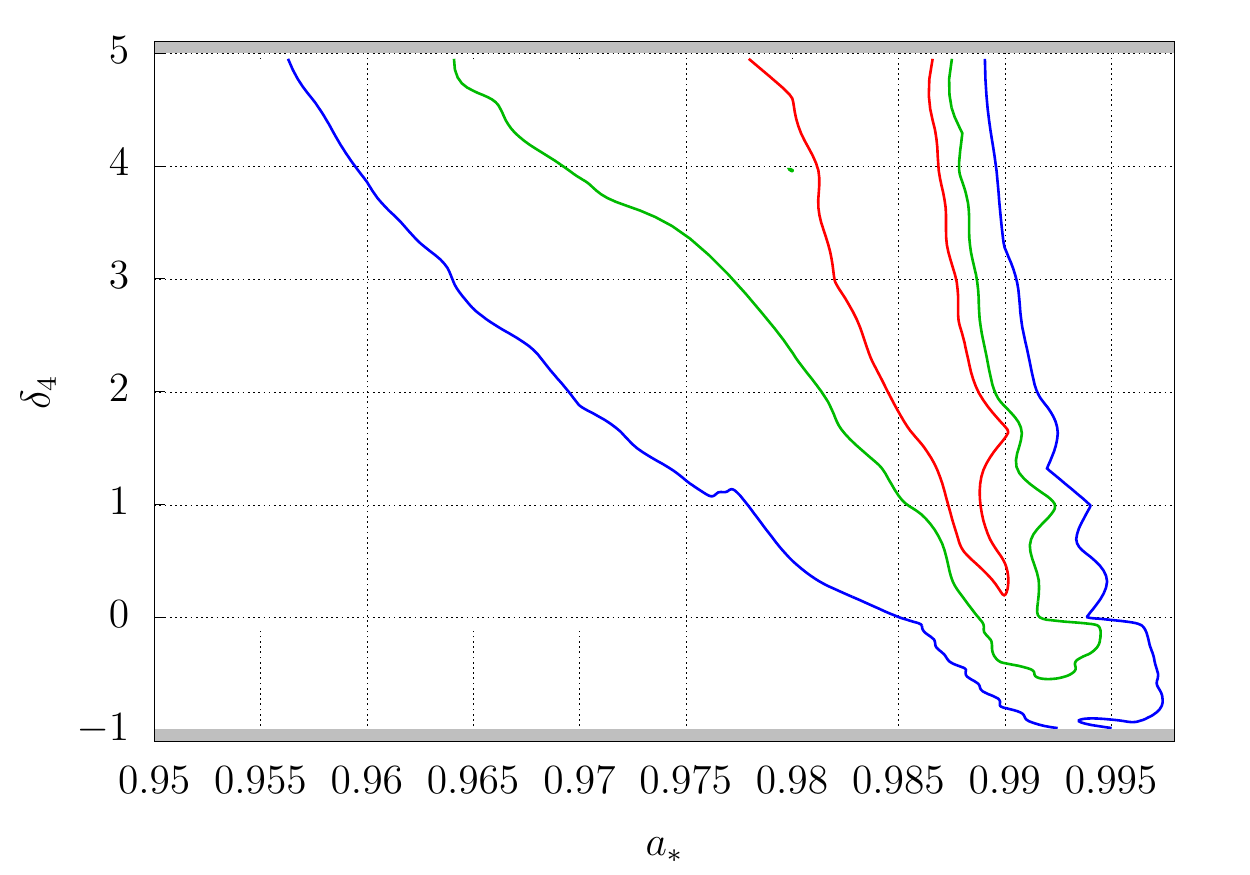}
      \includegraphics[width=\columnwidth]{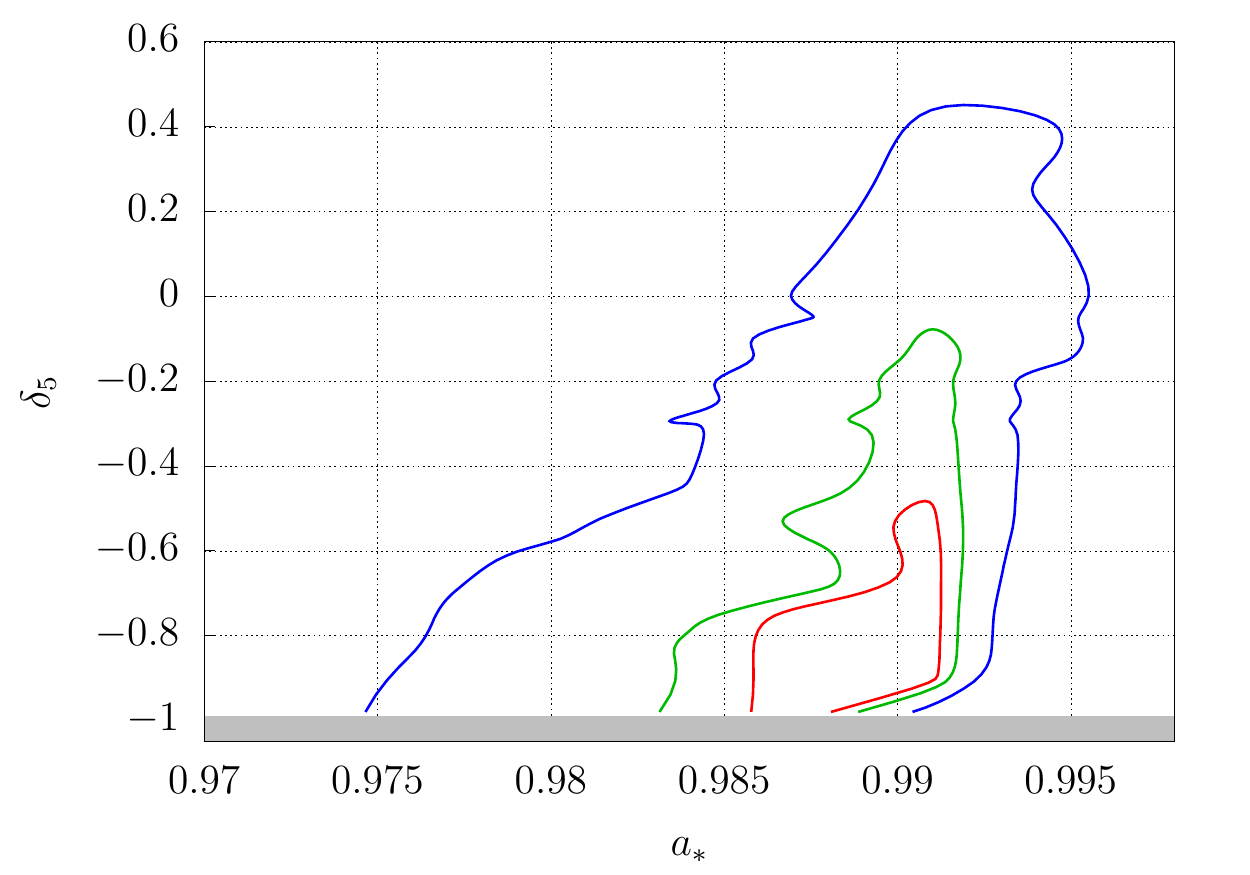}
     \includegraphics[width=\columnwidth]{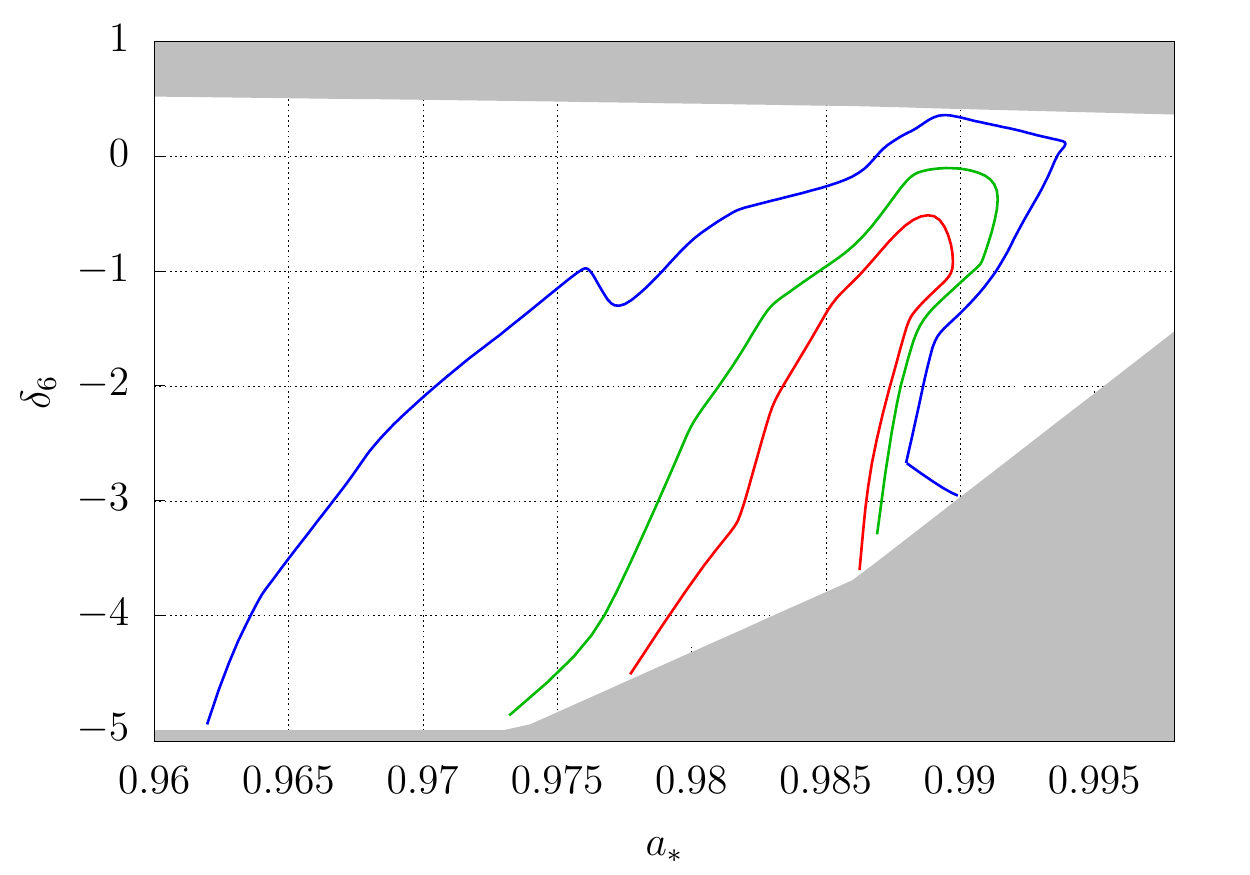}
    \caption{Degeneracy contours of the spin parameter $a_*$ and the KRZ deformation parameter $\delta_i$, respectively, within Model 2 (see Sec.~\ref{s-red}). The red, green, and blue lines indicate, respectively, the 68\%, 90\%, and 99\% confidence level contours for two relevant parameters. The grey region indicates parameter space which is excluded (see Sec.~\ref{s-trf}). The Kerr solution lies at $\delta_i=0$ in each plot.}
    \label{fig:contours}
\end{figure*}

\ch{The primary aim of this analysis is to study the spacetime metric around the supermassive BH in Ark~564 and estimate the deformation parameters $\delta_i$. Tab.~\ref{tab:params} lists the best-fit value of each $\delta_i$ and the associated uncertainty (at $90\%$ CL). In particular,
\begin{align}
	\begin{aligned}
	-0.27 < & \,\delta_1 < 0.28 \, , \nonumber \\
	-0.37 < & \,\delta_2 < 0.22 \, , \nonumber \\
	-0.45 < & \,\delta_3 < 1.04 \, , \nonumber \\
	-0.55 < & \,\delta_4\, , \nonumber \\
	-1 < & \,\delta_5 < -0.08\, (0.45) \, , \nonumber \\
	-4.9 < & \,\delta_6 < -0.10\, (0.36) \, .
	\end{aligned}
\end{align}
As we saw in the qualitative discussion of Sec.~\ref{sec:qual}, different parameters affect the reflection spectrum differently and thus, using the reflection data of Ark 564, are constrained to different levels. Specifically, $\delta_1$ is strongly constrained on the positive side whereas on the negative side the bound comes from the spacetime regularity conditions of Sec.~\ref{s-metric}. $\delta_2$ is strongly constrained on the negative side (considering the best-fit value of $\delta_2$, $-0.25$) and moderately constrained on the positive side, while $\delta_3$ is constrained strongly on the negative side and almost unconstrained on the positive side. $\delta_4$ has almost no effect on the reflection spectrum and consequently is very poorly constrained, in fact the upper bound exceeds the grid size which is at $\delta_4=5$. The best-fits for $\delta_5$ and $\delta_6$ are both quite far from $0$, though both have large uncertainties with $\delta_5$ hitting the theoretical limit (based on regularity of spacetime) and $\delta_6$ very close to the computational limit (at $\delta_6=-5$). In both $\delta_5$ and $\delta_6$ cases, the Kerr solution is recovered at $99\%$ CL (shown in parentheses).}

\ch{The KRZ metric has been used to perform tests of GR with other experimental techniques as well. Gravitational waves, in particular, have been used extensively in recent years to test GR in the strong field regime. The spherically symmetric version of the KRZ metric~\cite{Rezzolla:2014mua} has been used in Ref.~\cite{Cardenas-Avendano:2019zxd} to put constraints on one of its deformation parameters, using data from the first two observing runs of LIGO/VIRGO. The parameter analyzed in Ref.~\cite{Cardenas-Avendano:2019zxd}, $a_1$, is the same as $\delta_1$ of the present work. Since all the other parameters in that work are set to their GR value, the estimates obtained there can be faithfully compared with those obtained in the present work. Comparing the constraint on $\delta_1$ from above with Tab. 1 in Ref.~\cite{Cardenas-Avendano:2019zxd}, we see that the constraints of the present work are of the same order as Ref.~\cite{Cardenas-Avendano:2019zxd}: better than those obtained from the gravitational wave event GW170104 and worse than the other four events. It is important to remember though that the study in Ref.~\cite{Cardenas-Avendano:2019zxd} was done under the assumption of spherically symmetric BHs, which certainly will not be valid for all the gravitational wave events, and inclusion of spin will worsen the estimates there. Another recent work~\cite{PhysRevLett.125.141104} uses the results from the observation of the supermassive BH at the center of the M$87$ galaxy, reported by the Event Horizon Telescope collaboration~\cite{Akiyama:2019cqa}, to place constraints on the leading order strong-gravity deformation parameter, same as our $\delta_1$. The constraints obtained there are a few times weaker than those reported here. The constraints are summarized in Tab.~\ref{tab:delta1constraints}.}  
\begin{table*}[t]
\renewcommand{\arraystretch}{1.4}
 \begin{center}
  \begin{tabular}{>{\centering\arraybackslash}m{2.5cm} >{\centering\arraybackslash}m{2.5cm} >{\centering\arraybackslash}m{2.5cm} >{\centering\arraybackslash}m{5cm} }
     \hline
   \hline
   \textbf{Lower bound} & \textbf{Upper bound} & \textbf{Source} & \textbf{Technique}\\\hline
  --0.16 & +0.17 & GW151226 & Gravitational wave inspiral~\cite{Cardenas-Avendano:2019zxd}\\\hline
  --0.27 & +0.28 & Ark~564 & This work\\\hline
  --1.2 & +1.3 & M$87^*$ & BH shadow size~\cite{PhysRevLett.125.141104}\\
     \hline
   \hline
  \end{tabular}
    \caption{\ch{Lower and upper bounds on the KRZ deformation parameter $\delta_1$ from three different techniques. In the case of gravitational waves, the best constraints are reported, as obtained from the event GW191226. In the other two cases, only one source/observation has been used as of now to estimate the constraint.}}
    \label{tab:delta1constraints}
 \end{center}
\end{table*}

\ch{It is well known that model parameters can be degenerate, so while estimating them one needs to be careful in handling these degeneracies. We take care of this by marginalizing over all the free model parameters, and the uncertainties reported in Tab.~\ref{tab:params} are therefore robust statistically (systematic uncertainties are a different story, see the discussion later). 
In particular, there exists a strong degeneracy between $a_*$ and the deformation parameters, since both appear in the metric and, consequently, affect the spacetime in a somewhat similar way. Therefore, any test of GR has to contend with the fact that simultaneous measurement of both spin and non-Kerr parameters will have an intrinsic degeneracy, resulting in weaker constraints on each parameter (than if the BH was not spinning or if the non-Kerr parameters were identically zero). Fig.~\ref{fig:contours} shows this for the fits made with the Ark~564 data discussed above\footnote{Specifically, after obtaining the best-fit model in each $\delta_i$ case, we use Xspec's steppar command over two dimensions, $a_*$ and $\delta_i$, which steps through a series of pairs of ($a_*,\delta_i$) values, finds a new fit at each step by marginalizing over all other free parameters, and reports the $\chi^2$ at each step. The contours in the plots represent the change in $\chi^2$ at each step relative to the $\chi^2$ of the original best-fit.}, with each panel showing the relative degeneracy between $a_*$ and one of the $\delta_i$. The red, green, and blue lines indicate, respectively, the 68\%, 90\%, and 99\% confidence level contours. Generically, we can say that the constraints on each $\delta_i$ would be very different if the degeneracy with spin was not taken into account. For instance, we see in the panel with $\delta_6$ (bottom right panel) that at fixed spin ($a_*=0.99$), $\delta_6=-2$ is distinct from $\delta_6=0$, excluded at more than $99\%$ CL. This agrees with the qualitative analysis of Sec.~\ref{sec:qual}. But the analysis there was limited to constant BH spin. When the spin is allowed to vary, a larger range of $\delta_6$ ($\delta_6=-4$ at $a_*=0.97$) becomes degenerate.}
\ch{The degeneracy manifests in another way in the contours between $a_*$ and $\delta_1$ (top left panel) and $a_*$ and $\delta_2$ (top right panel), respectively. As discussed in Sec.~\ref{sec:qual}, the ISCO contour can be used as a proxy for the degeneracy, especially near the Kerr solution. In both the panels, we find that the degeneracy contours follow the constant ISCO contours close to the Kerr solution (see Fig.~\ref{fig:iscocon}). For larger deviations away from Kerr, the spectrum (which is affected by several things and not just the ISCO location) becomes distinct, and the ISCO ceases to be the driver of degeneracy.}


\ch{One of the aspects beyond the scope of this work that is important for parameter estimation is systematic uncertainties. They arise from the fact that the analysis makes a series of assumptions. The model, for example, assumes a thin disk described by the Novikov-Thorne model~\cite{Novikov1973}. For astrophysical systems, this is a valid assumption only when the accretion rate is not too high (nor too low)~\cite{Penna:2010hu,Kulkarni:2011cy}. Investigations with thick disks have shown that this assumption can potentially cause systematic error in spin up to $0.2$~\cite{Taylor:2017jep} and $0.4$~\cite{Riaz:2019kat}, and more than $1$ for deformation parameters~\cite{Riaz:2019bkv}. Another assumption, that particles move in circular orbits on the disk can also lead to systematic errors~\cite{Tripathi:2020wfi}. Higher order effects in the context of disk reflection, neglected in the current models, can be important in some cases~\cite{Zhu:2012vf,Zhou:2019dfw,Cardenas-Avendano:2020xtw}. The model is limited to only one non-zero deformation parameter. Degeneracy among the various deformation parameters is, therefore, beyond the scope of the model. While this is fine for verifying GR, mapping to BH metrics from alternative theories of gravity will certainly involve multiple deformation parameters. In this context, a recent study based on gravitational wave ringdown~\cite{Volkel:2020daa} has shown the severe weakening of constraints on each when multiple deformation parameters are set free.}

\ch{We emphasize that this work is an initial exploration with several aspects to be explored in future. These include, but are not limited to, studies involving more parameters (e.g., estimating $\epsilon_0$ parameter, see Eq.~\ref{eq:metric}), estimating multiple deformation parameters simultaneously (e.g., as done in~\cite{Volkel:2020daa}), employing more sophisticated disk geometry (e.g., as done in~\cite{Abdikamalov:2020oci}), studying more astrophysical sources (e.g., MCG-06-30-15~\cite{Fabian:2002gj,Marinucci:2014ita,Tripathi:2018lhx}, GRS 1915+105~\cite{McClintock:2006xd,Miller:2013rca,Zhang:2019ldz}, GX 339-4~\cite{Reis:2008ja,Garcia:2015uta,Wang-Ji:2018ssh}), and exploring potential synergy with other observational techniques~\cite{Cardenas-Avendano:2019zxd,Nampalliwar:2020asd,Volkel:2020daa}.} 


{\bf Acknowledgments --}
\ch{We thank Alejandro Cardenas-Avendano for pointing out the issue with matching the metric to the Kerr solution, and Roman Konoplya and Alexander Zhidenko for providing the correct metric components. We also thank the anonymous referee whose valuable feedback helped us improve the paper.} S.N. acknowledges support from the Excellence Initiative at Eberhard-Karls Universit\"at T\"ubingen. S.N. and J.A.G. also acknowledge support from the Alexander von Humboldt Foundation. A.B.A. acknowledges the support from the Shanghai Government Scholarship (SGS). A.T. acknowledges support from the China Scholarship Council (CSC), Grant No.~2016GXZR89. \ch{The authors acknowledge support by the High Performance and Cloud Computing Group at the Zentrum f\"{u}r Datenverarbeitung of the University of T\"{u}bingen, the state of Baden-W\"{u}rttemberg through bwHPC and the German Research Foundation (DFG) through grant no.~INST 37/935-1 FUGG.} This work was supported by the National Natural Science Foundation of China (NSFC), Grant No.~U1531117, and Fudan University, Grant No.~IDH1512060.


\bibliography{references}
\end{document}